\documentclass[lettersize,journal,twoside]{IEEEtran}
\usepackage{amsmath}
\usepackage{amsfonts}
\usepackage{etoolbox}
\patchcmd{\subequations}{}%
{}{}{}

\usepackage[ruled, algonl, vlined]{algorithm2e}
\SetAlFnt{\footnotesize}
\SetAlCapFnt{\footnotesize}
\SetAlCapNameFnt{\footnotesize}

\usepackage{array}
\usepackage{booktabs}
\usepackage[caption=false,font=normalsize,labelfont=sf,textfont=sf]{subfig}
\usepackage{cite}
\usepackage{enumitem}
\usepackage{float}
\usepackage{graphicx}
\usepackage{multirow}
\usepackage{nomencl}
\usepackage{stfloats}
\usepackage{setspace}
\usepackage{textcomp}
\usepackage{tabularx}
\usepackage{url}
\usepackage{verbatim}
\usepackage{makecell}
\usepackage{dcolumn}

\usepackage[colorlinks=true,
linkcolor=blue,
citecolor=blue,
urlcolor=blue]{hyperref}

\newcommand{\myLarge}{\fontsize{16.3pt}{17pt}\selectfont}

\begin{document}
\allowdisplaybreaks

\title{\myLarge
DRL-Based Medium-Term Planning of Renewable-Integrated Self-Scheduling Cascaded Hydropower to Guide Wholesale Market Participation}
\vspace{-8mm}
\author{Xianbang Chen,~\IEEEmembership{Student Member,~IEEE,}~Yikui Liu,~\IEEEmembership{Member,~IEEE,},~Neng Fan,~Lei Wu,~\IEEEmembership{Fellow,~IEEE}
\vspace{-10mm}

\thanks{This work is supported in part by the U.S. Department of Energy’s Office of Energy Efficiency and Renewable Energy (EERE) under the Water Power Technologies Office Award Number DE-EE0008944. The views expressed herein do not necessarily represent the views of the U.S. Department of Energy or the United States Government.

X. Chen and L. Wu are with the ECE Department, Stevens Institute of Technology, Hoboken, NJ, 07030 USA. (Email: xchen130; lei.wu@stevens.edu). Y. Liu is with the Electrical Engineering Department, Sichuan University, Chengdu, 610017 China. N. Fan is with the Department of Systems and Industrial Engineering, University of Arizona, Tucson, AZ, 85721 USA.}
}


\maketitle
\begin{abstract}
For self-scheduling cascaded hydropower (S-CHP) facilities, medium-term planning is a critical step that coordinates water availability over the medium-term horizon, providing water usage guidance for their short-term operations in wholesale market participation. Typically, medium-term planning strategies (e.g., reservoir storage targets at the end of each short-term period) are determined by either optimization methods or rules of thumb. However, with the integration of variable renewable energy sources (VRESs), optimization-based methods suffer from deviations between the anticipated and actual reservoir storage, while rules of thumb could be financially conservative, thereby compromising short-term operating profitability in wholesale market participation. This paper presents a deep reinforcement learning (DRL)-based framework to derive medium-term planning policies for VRES-integrated S-CHPs (VS-CHPs), which can leverage contextual information underneath individual short-term periods and train planning policies by their induced short-term operating profits in wholesale market participation. The proposed DRL-based framework offers two practical merits. First, its planning strategies consider both seasonal requirements of reservoir storage and needs for short-term operating profits. Second, it adopts a multi-parametric programming-based strategy to accelerate the expensive training process associated with multi-step short-term operations. Finally, the DRL-based framework is evaluated on a real-world VS-CHP, demonstrating its advantages over current practice.
\end{abstract}

\vspace{-2mm}

\vspace{-4mm}
\section*{Nomenclature}
\vspace{-2mm}
\addcontentsline{toc}{section}{Nomenclature}
\begin{spacing}{1}
\subsection*{Sets and Indices}
\vspace{-1mm}
\begin{IEEEdescription}[\IEEEusemathlabelsep \IEEEsetlabelwidth{$\mspace{36mu}$} \setlength{\IEEElabelindent}{0pt}]
\item[$\mathcal{B}_{r}$]
Active set of critical region (CR) $r$.

\item[$\mathcal{I}_{n}$]
Set of hydropower units on reservoir $n$, indexed by $i$.

\item[$\mathcal{J}/j_{t}$]
Set/index of sub-hourly intervals (SIs) in hour $t$. $j_{t}=1,...,J$ and $J=\text{12}$.

\item[$\mathcal{K}/k$]
Set/index of operation days in a water year. $k\in \mathcal{K}=\{1,...,K\}$ and $K=\text{366/365}$ in a leap/non-leap year.

\item[$\mathcal{N}/n$]
Set/index of reservoirs. $n\in \mathcal{N} = \{1,..., N\}$.

\item[$\bar{\mathcal{N}}_{n}/m$]
Set/index of direct upstream reservoirs of reservoir $n$.

\item[$\mathcal{R}/r, r^{\prime}$]
Set/indices of CRs. $r, r^{\prime} \in \mathcal{R}=\{1,..., R\}$.

\item[$\mathcal{T}/t$]
Set/index of hours. $t\in \mathcal{T} = \{1,..., T\}$ and $T=\text{24}$.

\end{IEEEdescription}

\vspace{-5mm}
\subsection*{Scenario-Independent Decision Variables}
\vspace{-1mm}
\begin{IEEEdescription}[\IEEEusemathlabelsep \IEEEsetlabelwidth{$\mspace{30mu}$} \setlength{\IEEElabelindent}{0pt}]
\item[$\boldsymbol{a}_{k}$]
Vector of medium-term planning strategies for day $k$.

\item[$G_t$]
Gross revenue of hour $t$. [\$]

\item[$P_{t}^{\text{sch}}$]
Self-scheduling power generation plan of hour $t$. [MW]

\item[$\bar{[\,\cdot\,]}$]
Indicating variables with known values.

\item[${[\,\cdot\,]}^{\text{RT}}$]
Indicating real-time operation variables.

\item[${[\,\cdot\,]}^{\star}$]
Indicating the optimal solution to a variable.

\end{IEEEdescription}

\vspace{-2mm}
\subsection*{Decision Variables Associated With Scenario $w$}
\vspace{-1mm}
\begin{IEEEdescription}[\IEEEusemathlabelsep \IEEEsetlabelwidth{$\mspace{30mu}$} \setlength{\IEEElabelindent}{0pt}]

\item[$C_{wt}$]
Penalty for under-generation in hour $t$. [\$]

\item[$D_{wnit}$]
Discharge rate of unit $i$ on reservoir $n$ in hour $t$. [$\text{m}^{\text{3}}$/s]

\item[$I_{wnit}$]
On/off status of unit $i$ on reservoir $n$ in hour $t$.

\item[$P_{wnit}$]
Scheduled power of unit $i$ on reservoir $n$ in hour $t$. [MW]

\item[$P_{wj_{t}}^{\text{act}}$]
Actual delivered power in SI $j_{t}$. [MW]

\item[$P_{wt}^{\text{vr}}$]
Scheduled power of VRES in hour $t$. [MW]

\item[$P_{wj_{t}}^{\Delta}$]
Under-generated power in SI $j_{t}$. [MW]

\item[$V_{wnt}$]
Storage volume of reservoir $n$ at the beginning of hour $t$. [$\text{Mm}^{\text{3}}$]

\item[$W_{wnt}^{\text{ws}}$]
Water spillage of reservoir $n$ in hour $t$. [$\text{Mm}^{\text{3}}$]

\item[$W_{wnt}^{\text{i/o}}$]
Water scheduled to flow into/out of reservoir $n$ in hour $t$. [$\text{Mm}^{\text{3}}$]
\end{IEEEdescription}

\vspace{-5mm}
\subsection*{Parameters}
\vspace{-1mm}
\begin{IEEEdescription}[\IEEEusemathlabelsep \IEEEsetlabelwidth{$\mspace{30mu}$} \setlength{\IEEElabelindent}{0pt}]

\item[$C^{\Delta}_{j_{t}}$]
Penalty coefficient for power under-generation. [\$/MW]

\item[$C^{\text{re}}$]
Reward coefficient for reserving water. [\$/$\text{Mm}^{\text{3}}$]

\item[$D_{ni}^{\text{M/m}}$]
Max/min discharge rate of unit $i$ on reservoir $n$. [$\text{m}^{\text{3}}$/s]

\item[$P_{ni}^{\text{M/m}}$]
Max/min power of unit $i$ on reservoir $n$. [MW]

\item[$\hat{P}^{\text{vr}}_{t}$]
Predicted VRES availability in hour $t$. [MW]

\item[$\tilde{P}^{\text{vr}}_{j_{t}}$]
Actual VRES availability in SI $j_{t}$. [MW]

\item[$V_{n}^{\text{M/m}}$]
Max/min storage volume of reservoir $n$. [$\text{Mm}^{\text{3}}$]

\item[$\hat{V}_{n}^{\text{end}}$]
Target end-of-day storage of reservoir $n$. [$\text{Mm}^{\text{3}}$]

\item[$\tilde{V}_{n}^{\text{end}}$]
Actual end-of-day storage of reservoir $n$. [$\text{Mm}^{\text{3}}$]

\item[$V_{n}^{\text{ini}}$]
Initial storage of reservoir $n$. [$\text{Mm}^{\text{3}}$]

\item[$\hat{W}_{nt}$]
Predicted water inflow of reservoir $n$ in hour $t$. [$\text{Mm}^{\text{3}}$]

\item[$\tilde{W}_{nj_{t}}$]
Actual water inflow of reservoir $n$ in SI $j_{t}$. [$\text{Mm}^{\text{3}}$]

\item[$\alpha, \beta$]
Coefficients converting $\text{m}^{\text{3}}$/s to $\text{Mm}^{\text{3}}$. [s]

\item[$\sigma^{\text{H}}$]
Duration of 1 hour. [hour]

\item[$\lambda_{t}$]
Locational marginal price of hour $t$. [\$/MWh]

\item[$\hat{\nu}_{n}$]
Estimated water value of $V_{n,T+1}$. [\$/$\text{Mm}^{\text{3}}$]

\item[$\Theta$] 
Entire feasible region of parameters.

\item[$\Theta_{r}^{\text{CR}}$]
CR $r$, where $\Theta^{\text{CR}}_{r} \subseteq \Theta$.

\end{IEEEdescription}
\end{spacing}

  \setlength{\abovedisplayskip}{1pt}
  \setlength{\belowdisplayskip}{\abovedisplayskip}
  \setlength{\abovedisplayshortskip}{0pt}
  \setlength{\belowdisplayshortskip}{1pt}

   \vspace{-3mm}
\section{Introduction}
   \vspace{-1mm}
\subsection{Background}
\begin{figure}[tb]
	\centering
		\includegraphics[width=\columnwidth]{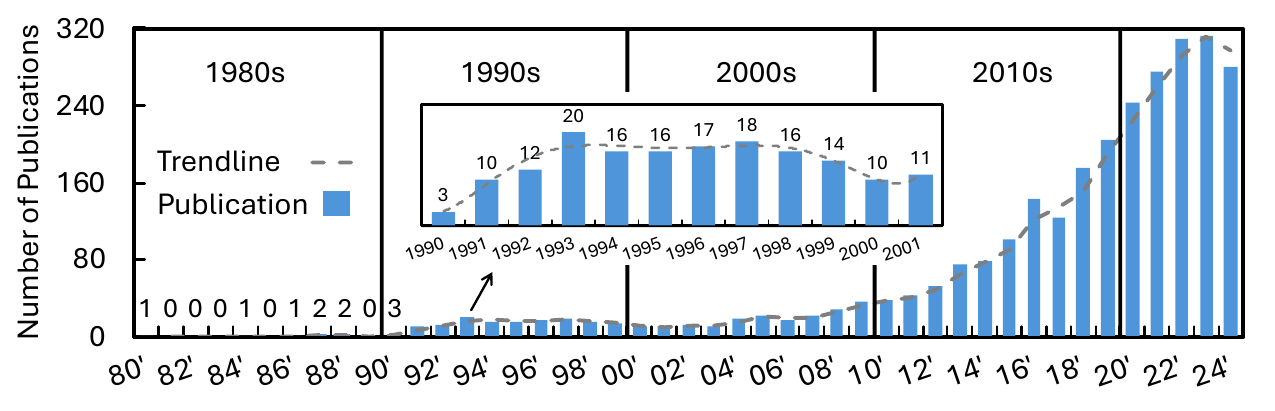}
  \vspace{-7mm}
	\caption{Web of Science Core Collection query results (as of 2023/12/31): TS=((“hydro” OR “hydropower” OR “hydroelectricity”) AND (“power system” OR “electricity system” OR “power grid” OR “electricity grid” OR “electricity market” OR “energy trading”)) AND DT=(Article).}\label{fig01}
\vspace{-5mm}
\end{figure}

\IEEEPARstart{H}{ydropower}, as the oldest renewable energy technology, contributes to approximately 28.7\% of U.S. renewable electricity generation in 2024 \cite{hydrowires}. As shown in Fig.~\ref{fig01}, hydropower research related to power systems reached a modest peak in the mid-1990s and then declined slightly in the early 2000s. From 2010 onwards, interest in such research surged significantly, largely driven by the deeper integration of variable renewable energy sources (VRESs) such as wind and solar \cite{Arild_IET_Hydro_Wind_Origin}. Indeed, integrating VRESs into power system operations necessitates a more effective utilization of the flexibility and rapid responsiveness of hydropower \cite{Arild_TPS_Optimal_Hydropower_Maintenance}. As a result, many U.S. hydropower fleets are currently transforming into a new role as critical integrators of VRESs.

Significant efforts are being made in the U.S. to support this transformation. One notable example is the HydroWIRES Initiative \cite{hydrowires} launched in 2019 by the U.S. Department of Energy's Water Power Technologies Office to investigate new hydropower operation strategies with enhanced flexibility and value streams. Efforts similar to this initiative have been continuously incentivizing existing hydropower fleets to evolve into their hybrid hydro-VRES (HHVRES) counterparts. For instance, Portland General Electric (PGE), a self-scheduling \cite{DT_SelfScheduling} participant\footnote{Self-scheduling producers are market participants who design generation plans themselves and submit them to the regional transmission operator (RTO)/independent system operator (ISO), instead of letting the RTO/ISO optimize their assets from the entire system's point of view.} in the Western Energy Imbalance Market (WEIM) of California ISO (CAISO), has begun exploring its potential to transform into an HHVRE. However, this transformation also introduced greater complexity in hydropower scheduling tasks across multiple time scales, driving the need for more advanced scheduling frameworks.

In order to facilitate the transformation of self-scheduling cascaded hydropower (S-CHP) into HHVRESs, this paper studies effective medium-term planning strategies to guide VRES-integrated S-CHP (VS-CHP) with enhanced water usage for short-term operations in wholesale market participation, using PGE as a real-world case.

\vspace{-3mm}
\subsection{Related Works and Challenges}\label{Gap}
\begin{figure}[b]
\centering
    \vspace{-5mm}
		\includegraphics[width=0.95\columnwidth]{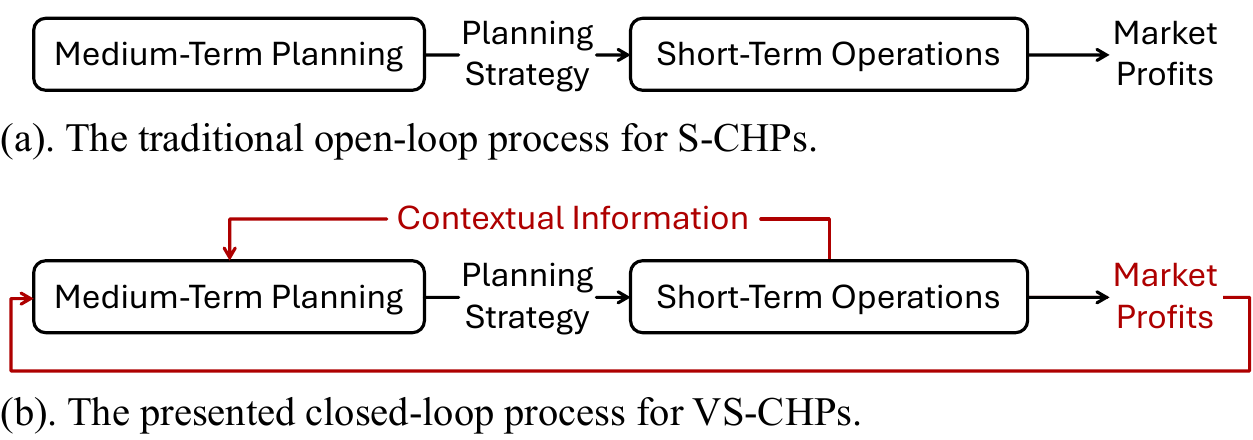}
  \vspace{-3mm}
	\caption{Comparisons of different scheduling processes.}\label{fig02}
\end{figure}

An S-CHP operator needs to coordinate multiple tasks across different time scales to design its generation plans. Specifically, as sketched in Fig.~\ref{fig02}(a), medium-term planning is first conducted weeks ahead \cite{Arild_TSTE_Optimal_Midterm, Arild_SDDiP, Arild_IET_Assessing_Profitability, Arild_J_Water_Resour_Plan_Manag_SDDP}, and then the planning strategy is used to guide multi-step short-term operations \cite{Arild_TPS_Short_Term_Convex_Relaxations} in wholesale market participation. The ultimate goal is to maximize profits from the wholesale market.

Traditional medium-term S-CHP planning methods mainly involve multi-stage stochastic optimization \cite{Arild_TSTE_Optimal_Midterm, Arild_SDDiP, Arild_IET_Assessing_Profitability, Arild_J_Water_Resour_Plan_Manag_SDDP} and rules of thumb \cite{DT_rule_of_thumb}, which could be ineffective under the transformation towards VS-CHPs. Specifically, optimization-based planning results can be inefficient (or even physically infeasible) for short-term operations, as actual water storage of reservoirs may significantly deviate from the planned optimal volumes; the rule-of-thumb method, e.g., maintaining equal end-of-day and beginning-of-day storage \cite{DT_rule_of_thumb}, does not fully leverage the VRES context and can be financially conservative. The limitations of these methods stem from two issues: \textit{i)} they do not fully use context relevant to short-term operations, such as short-term VRES predictions and reservoirs status, and \textit{ii)} they are myopic regarding their impact on short-term operations because of the open-loop relationship between medium-term planning and short-term operations, as shown in Fig.~\ref{fig02}(a).

Regarding these limitations, deep reinforcement learning (DRL), recognized for its closed-loop nature \cite{yufan_DRL_3_closed_loop} and remarkable end-to-end learning capability \cite{yufan_DRL_1_contextual} in complex environments, emerges as a promising alternative. DRL has been applied to VS-CHP scheduling \cite{DRL-hydro-Jingxian-Yang-2, DRL-hydro-Huang-Qin, DRL-hydro-Jiang, DRL-hydro-Zeng} and similar tasks \cite{DRL-hydro-Longoria, DRL-Bertrand, QT_SAC_1, QT_SAC_2, QT_SAC_3}. However, existing DRL applications mainly focus on short-term VS-CHP operation policies, neglecting medium-term water usage coordination that otherwise could further boost short-term operations with efficient water usage. To this end, this paper applies DRL to medium-term VS-CHP planning while considering the planning impact on the short-term operating profits in the wholesale market, thereby integrating medium-term planning with short-term operations in a closed-loop fashion, as shown in Fig.~\ref{fig02}(b). It will provide VS-CHP operators with enhanced water usage guidance and great flexibility for short-term operations while avoiding unfavorable market profit outcomes.

Nevertheless, applying DRL to the medium-term planning is non-trivial due to two challenges: \textit{i)}
besides pursuing market profit, hydropower facilities must adapt reservoir storage to seasonality, e.g., refilling reservoirs before dry seasons and restoring storage levels by the end of the water year, to meet stringent water regulation requirements. However, existing DRL-based studies (e.g., \cite{DRL-hydro-Jingxian-Yang-2, DRL-hydro-Huang-Qin, DRL-hydro-Jiang, DRL-hydro-Zeng}) mainly focus on maximizing short-term profits and do not capture such seasonal adaptability; and \textit{ii)} training medium-term policies in an environment that closely mimics multi-step short-time operations in wholesale market participation is essential for ensuring the out-of-sample performance. However, the training process can be computationally taxing because each training step involves solving hundreds of operation models on a rolling basis to evaluate the performance in real-time (RT) market participation. To the authors' knowledge, only a few works (e.g., \cite{DRL-hydro-Longoria, DRL-Bertrand}) have considered this challenge, but through simplified optimization models that cannot adequately capture the essential operation characteristics of VS-CHPs.

\vspace{-3mm}
\subsection{Proposed Works}
To resolve the challenges from the literature review, this paper presents a computationally efficient DRL-based medium-term VS-CHP planning framework. It provides enhanced water usage guidance for short-term operations to achieve promising profits in wholesale market participation while maintaining seasonal adaptability. In summary, The presented framework offers the following two advantages:

\begin{itemize}[noitemsep, topsep=0pt, parsep=0pt, partopsep=0pt, leftmargin=*, wide = 0pt]
\item
Compared to existing medium-term planning studies (e.g., \cite{Arild_TSTE_Optimal_Midterm, Arild_SDDiP, Arild_IET_Assessing_Profitability, Arild_J_Water_Resour_Plan_Manag_SDDP}), the presented framework stands out with its closed-loop performance-driven evaluation \cite{yufan_DRL_3_closed_loop} and context-aware decision-making \cite{yufan_DRL_1_contextual}. As shown in Fig.~\ref{fig02}(b), the training process uses short-term contextual information to feed medium-term planning policies, ensuring that the derived planning strategies are timely and relevant. Moreover, the induced market profit is fed back to quantify planning strategies and update planning policies for further boosting market profits in short-term operations;

\item
Compared to related DRL applications \cite{DRL-hydro-Jingxian-Yang-2, DRL-hydro-Huang-Qin, DRL-hydro-Jiang, DRL-hydro-Zeng, DRL-hydro-Longoria, DRL-Bertrand}, the presented framework offers seasonal adaptability and efficiently mimics the multi-step short-term operations in wholesale market participation. Specifically, inspired by the game Space Impact \cite{SpaceImpact}, an expertise-based mechanism is incorporated into the training process to derive seasonality-aware policies. To handle the computational issue, an acceleration approach based on multi-parametric programming (MPP) \cite{mp_yufan2, mp_yufan3, inverseTrick, redundant} is deployed in the training process.
\end{itemize}

The remainder of the paper is organized as follows: Section~\ref{Sec2} introduces the DRL-based medium-term planning framework; Section~\ref{Sec3} details the enhancements in the DRL training to achieve seasonal adaptivity and computational tractability; Section~\ref{Sec4} presents numerical testing results on the PGE study; Section~\ref{Sec5} concludes the paper.

\vspace{-1mm}
\section{DRL-Based Medium-Term Planning Framework}\label{Sec2}
\subsection{Scheduling Timeline for VS-CHPs in Wholesale Markets}
\begin{figure}[b]
\vspace{-5mm}
	\centering
		\includegraphics[width=\columnwidth]{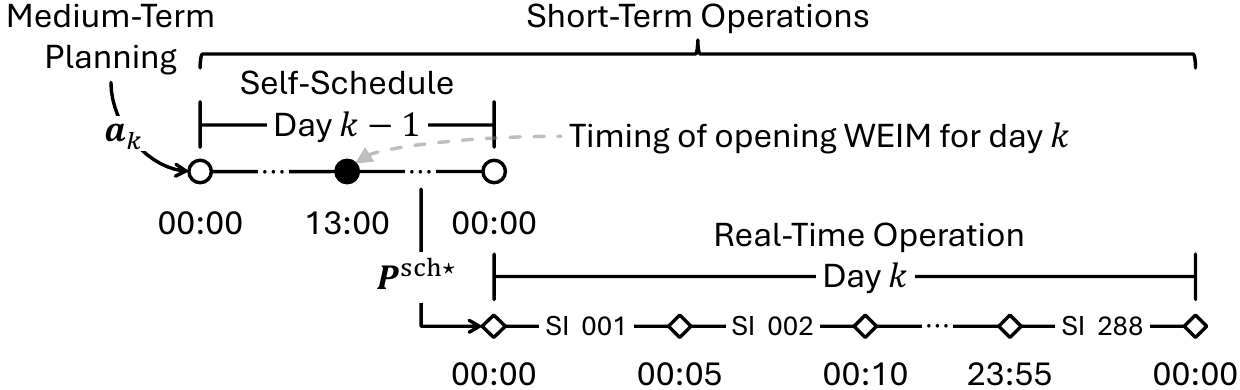}
  \vspace{-5mm}
	\caption{A typical timeline for VS-CHPs in WEIM participation of day $k$.}\label{fig03}
\end{figure}

VS-CHPs, like the one operated by PGE, conduct multiple operation tasks across different time scales to participate in CAISO's WEIM, as shown in Fig.~\ref{fig03}. First, medium-term planning is conducted weeks in advance to provide water usage guidance for day $k$, denoted as $\boldsymbol{a}_k$, which can take two forms: \textit{i)} end-of-day storage volumes \cite{Arild_TSTE_Optimal_Midterm}, i.e., the water volumes that individual reservoirs should reach by the end of day $k$; and \textit{ii)} water values \cite{Arild_IET_Assessing_Profitability} in reservoirs by the end of day $k$, i.e., market revenues that an incremental unit of water stored in individual reservoirs can potentially yield in the future.

Before day $k$, the VS-CHP determines its generation plan (denoted as $\boldsymbol{P}^{\text{sch}}$) for day $k$ by solving a mixed-integer linear programming (MILP)-based self-scheduling model. The compact form of the self-scheduling model is shown in \eqref{HModelA}, with medium-term planning strategy $\boldsymbol{a}_k$ and short-term context (i.e., locational marginal prices (LMPs) and VRES predictions) as input. Here, vector $\boldsymbol{h}$ and matrices $\boldsymbol{G}$ and $\boldsymbol{H}$ are known parameters; vectors $\boldsymbol{g}(\boldsymbol{a}_{k})$ and $\boldsymbol{l}(\boldsymbol{a}_{k})$ depend on $\boldsymbol{a}_k$; continuous variables $\boldsymbol{x}$ describe hourly storage levels of individual reservoirs as well as hourly water discharge rates and generation plans of individual hydro units; binary variables $\boldsymbol{y}$ denote hourly on/off statuses of individual hydro units. The objective function \eqref{HModelA:1} maximizes the net revenue in WEIM participation; \eqref{HModelA:2} involves prevalent hydropower scheduling constraints such as storage and discharge limits. The detailed formulation of \eqref{HModelA} is presented in Appendix~\ref{DAModel}.
\begin{subequations}\label{HModelA}
\begin{align}
&\textstyle{\max\nolimits_{\boldsymbol{x}, \boldsymbol{y}}}\,\,
\boldsymbol{g}(\boldsymbol{a}_{k})^{\top}\boldsymbol{x} + \boldsymbol{h}^{\top}\boldsymbol{y}                     \label{HModelA:1} \\
&\text{s.t. }\boldsymbol{G}\boldsymbol{x} + \boldsymbol{H}\boldsymbol{y} \leq \boldsymbol{l}(\boldsymbol{a}_{k});\,\boldsymbol{x} \in \mathbb{R}^{p}_{+},\, \boldsymbol{y} \in \{0, 1\}^{q};                       \label{HModelA:2}
\end{align}
\end{subequations}

\begin{figure}[tb]
	\centering
		\includegraphics[width=\columnwidth]{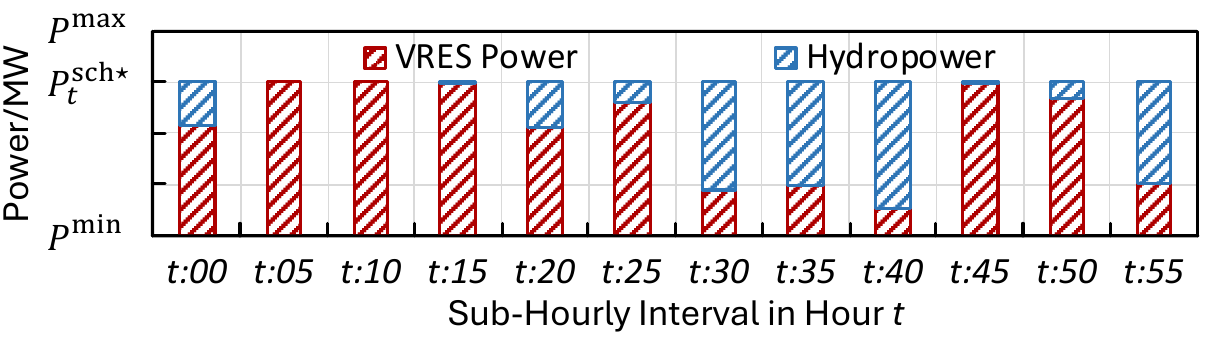}
  \vspace{-3mm}
	\caption{Real-time power mixture of HHVRES producers.}\label{fig04}
    \vspace{-6mm}
\end{figure}

Solving \eqref{HModelA} yields the optimal generation plan, $\boldsymbol{P}^{\text{sch}\star} = [{P}^{\text{sch}\star}_1,...,{P}^{\text{sch}\star}_T]$, which specifies hourly power levels that the VS-CHP will deliver on day $k$. This plan is then submitted to WEIM, and WEIM makes every effort to fully accept it \cite{CAISO2023}. Meanwhile, the VS-CHP is subject to strict delivery requirements that prohibit over-generation and impose penalties for under-generation. Thus, the VS-CHP generally leverages flexible hydropower to smooth out sub-hourly fluctuations of VRESs.  Specifically, in the RT stage of day $k$, the producers solve multiple MILP-based RT operation models in a rolling manner. The goal is to ensure that the power delivered in each 5-minute sub-hourly interval (SI) meets the generation plan accepted by WEIM, as shown in Fig.~\ref{fig04}.

The compact form of the RT operation problem is shown in \eqref{RTModelA},  which is built upon RT information on actual VRES availability, reservoir status, and water inflow. Here, vectors $\boldsymbol{c}$ and  $\boldsymbol{d}$ and matrices $\boldsymbol{A}$ and $\boldsymbol{E}$ are known parameters; the right-hand-side vector $\boldsymbol{b}(\boldsymbol{P}^{\text{sch}\star})$ depends on $\boldsymbol{P}^{\text{sch}\star}$; continuous variables $\boldsymbol{x}^{\text{RT}}$ describe sub-hourly storage levels of individual reservoirs as well as sub-hourly water discharge rates and generation plans of individual hydro units; binary variables $\boldsymbol{y}^{\text{RT}}$ represent on/off statuses of individual hydro units in each SI. In \eqref{RTModelA}, the objective function \eqref{RTModelA:1} minimizes the delivery deviation (denoted as $P^{\text{RT,}\Delta}_{j_{t}}$) in each SI $j_{t}$, and constraint \eqref{RTModelA:2} involves the prevalent hydropower scheduling constraints. The detailed formulation of \eqref{RTModelA} is presented in Appendix~\ref{DetailRT}.
\begin{subequations}\label{RTModelA}
\begin{flalign}
&\textstyle{\min\nolimits_{\boldsymbol{x}^{\text{RT}}, \boldsymbol{y}^{\text{RT}}}}\,\,\boldsymbol{c}^{\top}\boldsymbol{x}^{\text{\text{RT}}} + \boldsymbol{d}^{\top}\boldsymbol{y}^{\text{RT}}                                                            \mspace{-20mu}&\label{RTModelA:1} \\
&\text{s.t. }\mspace{-3mu} \boldsymbol{A}\boldsymbol{x}^{\text{RT}}\mspace{-3mu}+\mspace{-3mu}\boldsymbol{E}\boldsymbol{y}^{\text{RT}} \mspace{-3mu}\leq \mspace{-4mu}\boldsymbol{b}(\boldsymbol{P}^{\text{sch}\star});\,
\boldsymbol{x}^{\text{RT}}\mspace{-3mu} \in \mspace{-3mu}\mathbb{R}^{p^{\text{RT}}}_{+}\mspace{-2mu}, \mspace{-2mu}\boldsymbol{y}^{\text{RT}}\mspace{-3mu} \in \mspace{-1mu}\{0,\mspace{-2mu}1\}^{q^{\text{RT}}}\mspace{-3mu};           \mspace{-20mu} &\label{RTModelA:2}
\end{flalign}
\end{subequations}

The VS-CHP needs to solve 288 problems \eqref{RTModelA} corresponding to individual 5-minute SIs in day $k$. For each SI $j_{t}$, problem \eqref{RTModelA} has distinct right-hand-side vector $\boldsymbol{b}(\boldsymbol{P}^{\text{sch}\star})$, actual VRES availability $\tilde{P}^{\text{vr}}_{j_{t}}$, initial reservoir storage $\bar{\boldsymbol{V}}_{j_{t}}^{\smash{\raisebox{-0.5ex}{\scriptsize RT}}}$, scheduled inflow $\bar{\boldsymbol{W}}_{j_{t}}^{\smash{\raisebox{-0.5ex}{\scriptsize RT,i}}}$, and natural inflow $\tilde{\boldsymbol{W}}_{j_{t}}$.


 \vspace{-3mm}
\subsection{Formulate the Medium-Term Planning Problem as an MDP}
\vspace{-1mm}
To apply DRL, we first formulate the medium-term planning problem as a Markov decision process (MDP). This MDP is represented as a 4-tuple $(\mathcal{S}, \mathcal{A}, \mathcal{P}, \mathcal{R})$, where $\mathcal{S}$ is the set of states, $\mathcal{A}$ is the set of actions, $\mathcal{P}$ is the transition probability function, and $\mathcal{R}$ is the reward function.
In this MDP, each \textit{step} $k$ corresponds to one \textit{operation day}; each \textit{episode} represents one \textit{water year} which consists of $K$ operation days; \textit{action} $\boldsymbol{a}_{k}$ is the medium-term planning strategy; \textit{state} $\boldsymbol{s}_{k}$ refers to context describing the market and VS-CHP; the \textit{policy}, denoted as $\pi_{\theta}$, is a deep neural network (DNN) parameterized by $\theta$, which determines $\boldsymbol{a}_{k}$ based on $\boldsymbol{s}_{k}$, as expressed in \eqref{AgentandPolicy}. 
\begin{equation}\label{AgentandPolicy}
\textstyle{\boldsymbol{a}_k \sim \pi_{\theta}(\cdot | \boldsymbol{s}_k)}
\end{equation}

\begin{itemize}[noitemsep, topsep=0pt, parsep=0pt, partopsep=0pt, leftmargin=*, wide = 0pt]
\item
\textit{State}: 
The state $\boldsymbol{s}_{k}$ of operation day $k$ is defined in \eqref{State}, including day index $k$, inflow predictions $\hat{\boldsymbol{W}}_{k}$, VRES predictions $\hat{\boldsymbol{P}}_{k}^{\smash{\raisebox{-0.5ex}{\scriptsize vr}}}$, LMPs $\boldsymbol{\lambda}_{k}$, and initial storage volumes $\boldsymbol{V}^{\text{ini}}_{k}$.
\begin{equation}\label{State}
\textstyle{\boldsymbol{s}_{k}=
[k, \hat{\boldsymbol{W}}_{k}, \hat{\boldsymbol{P}}_{k}^{\smash{\raisebox{-0.5ex}{\scriptsize vr}}}, \boldsymbol{\lambda}_{k}, \boldsymbol{V}^{\text{ini}}_{k}]}
\end{equation}

\item
\textit{Action}: The action $\boldsymbol{a}_{k}$ is a $N$-dimensional vector representing the medium-term planning strategies. It can take the form of target end-of-day storage volume (denoted as $\hat{\boldsymbol{V}}_{k}^{\smash{\raisebox{-0.5ex}{\scriptsize end}}}$) or water values (denoted as $\hat{\boldsymbol{\nu}}_k$), as defined in \eqref{Action}.
\begin{equation}\label{Action}
\textstyle{\boldsymbol{a}_{k}= [\hat{V}_{k,1}^{\text{end}},...,\hat{V}_{k,N}^{\text{end}}]^{\top}
\text{ or }
\boldsymbol{a}_{k}=[\hat{\nu}_{k,1},...,\hat{\nu}_{k,N}  ]^{\top}}
\end{equation}

\item
\textit{Transition Probability Function}: After executing $\boldsymbol{a}_{k}$ in the self-scheduling model \eqref{HModelA}, the environment progresses through day $k$ following the timeline described in Fig.~\ref{fig03}. Upon reaching day $k+1$, the environment transitions from state $\boldsymbol{s}_{k}$ to state $\boldsymbol{s}_{k+1}$ probabilistically. The transition probability is described by function $\mathcal{P}(\boldsymbol{s}_{k+1}|\boldsymbol{a}_{k}, \boldsymbol{s}_{k}): \mathcal{S} \times \mathcal{A} \times \mathcal{S} \mapsto [0, 1]$, which is considered unknown and will be learned during training.

\item
\textit{Reward Function}: The reward function \eqref{Reward} calculates the net revenue as a result of executing the planning strategy $\boldsymbol{a}_{k}$ in state $\boldsymbol{s}_{k}$, where the first term calculates the gross revenue and the second term represents the under-generation penalty. The time-varying coefficient $C^{\Delta}_{j_{t}}$ follows a similar trend as RT LMPs to reflect dynamic penalties for under-generation under individual SI RT operations. By using the reward function \eqref{Reward} to train policy $\pi_{\theta}$, the goals of short-term operations and medium-term planning are closely aligned with pursuing the ultimate profit in WEIM participation.
\begin{equation}\label{Reward}
\textstyle{\mathcal{R}(\boldsymbol{s}_{k}, \boldsymbol{a}_{k})
= \sum\nolimits_{t=1}^{T}
[\sigma^{\text{H}}\lambda_{t} P^{\text{sch}\star}_{t}
- \sum\nolimits_{j_{t}=1}^{J}C^{\Delta}_{j_{t}} P^{\text{RT,}\Delta}_{j_{t}}]}
\end{equation}

\end{itemize}

\vspace{-2mm}
\subsection{Train Policy for Medium-Term Planning Decision-Making}
\vspace{-1mm}
Based on the MDP described above, state-of-the-art algorithms, e.g., soft actor-critic (SAC) \cite{QT_SAC_1, QT_SAC_2, QT_SAC_3}, can be used to train the policy and yield medium-term planning strategies. Note that off-the-shelf deep learning libraries, e.g., PyTorch and Gymnasium, enable the convenient implementation of various DRL algorithms within the framework. As shown in \eqref{Obj}, the training objective is to find a policy $\pi_{\theta}$ that maximizes the expected total net revenue in a $K$-day water year, where $\gamma_k$ is the discount factor for future rewards.
\begin{align}\label{Obj}
\theta^{\star} = \arg \textstyle{\max_{\theta}} \mathbb{E}_{\tau \sim \pi_{\theta}}
\textstyle{[\sum\nolimits_{k=1}^{K}\gamma_{k} \mathcal{R}(\boldsymbol{s}_{k}, \boldsymbol{a}_{k})]},\notag& \\
\text{where } \tau = (\boldsymbol{s}_{1}, \boldsymbol{a}_{1}, ..., \boldsymbol{s}_{K}, \boldsymbol{a}_{K})&
\end{align}

\begin{figure}[tb]
	\centering
		\includegraphics[width=\columnwidth]{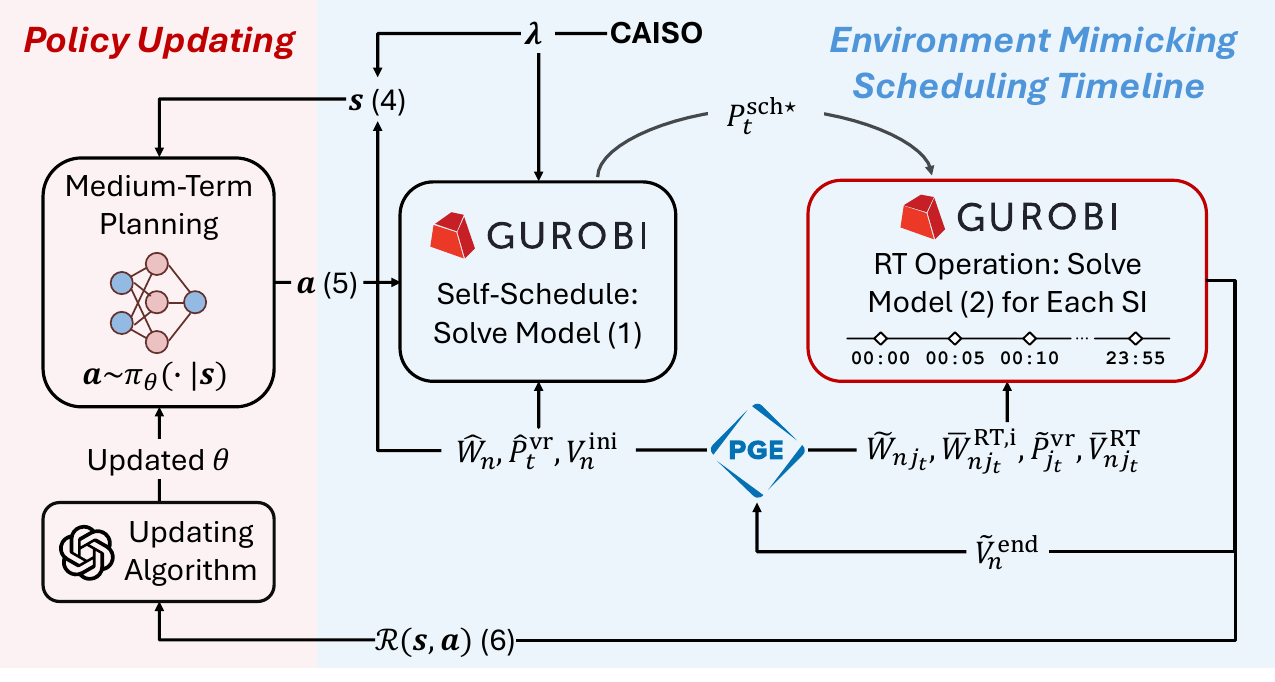}
	 \vspace{-8mm}
  \caption{The presented DRL-based framework.}\label{fig05}
  \vspace{-4mm}
\end{figure}

The training process, as shown in Fig.~\ref{fig05}, comprises two major parts: \textit{i)} a procedure (pink area) to update policy parameters $\theta$ and \textit{ii)} an environment (blue area) to simulate the VS-CHP's scheduling timeline in WEIM. Step $k$ of the training process begins in the pink area, where the medium-term planning strategy $\boldsymbol{a}_{k}$ is determined by the incumbent policy $\pi_{\theta}$ based on context $\boldsymbol{s}_{k}$. The process then moves into the blue area, where self-scheduling model \eqref{HModelA} is solved with input $\boldsymbol{a}_{k}$, followed by solving 288 RT models \eqref{RTModelA} on a rolling basis. Upon completing these, the revealed net revenue $\mathcal{R}_{k}$ is used to update policy parameters $\theta$, and the corresponding end-of-day storage $\tilde{\boldsymbol{V}}_{k}^{\smash{\raisebox{-0.5ex}{\scriptsize end}}}$ acts as the initial storage volume for step $(k+1)$. Afterward, step $(k+1)$ repeats the above process.

   \vspace{-1mm}
\section{Achieving Seasonal Adaptivity and\\ Training Acceleration}\label{Sec3}
As emphasized in Section~\ref{Gap}, applying DRL to medium-term VS-CHP planning is non-trivial due to the challenges in ensuring the seasonal adaptivity of end-of-day storage and achieving computational tractability of the policy training. Indeed, VS-CHPs prioritize seasonal adaptivity over net revenue due to stringent water regulation requirements, which, however, cannot be adequately captured by directly applying DRL.
In addition, as highlighted in the red box of Fig.~\ref{fig05}, each training step has to solve 288 RT operation models sequentially, which takes a few minutes even with Gurobi 11.0. Given that the training could require hundreds of thousands of steps, the training process will be extremely computationally taxing without proper acceleration methods.

\begin{figure}[tb]
	\centering
		\includegraphics[width=\columnwidth]{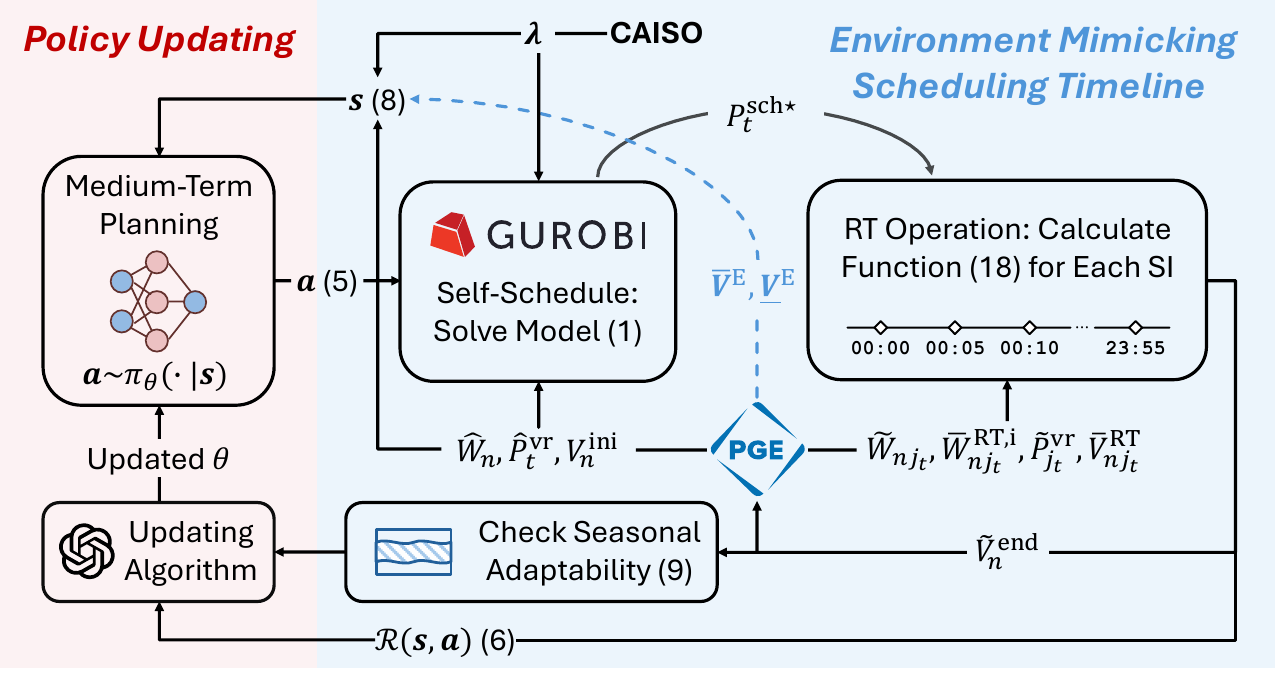}
  \vspace{-8mm}
	\caption{The presented DRL-based framework with enhancements.}\label{fig06}
 \vspace{-5mm}
\end{figure}

Regarding the two issues, two enhancements are presented to evolve the framework in Fig.~\ref{fig05} into an enhanced version in Fig.~\ref{fig06}. First, an expertise-based mechanism is introduced to enable the policy to generate end-of-day storage that satisfies the seasonal adaptivity requirements. Second, an MPP-based method is developed to replace the time-consuming RT optimizations with a computationally efficient process of logical operations, multiplications, and additions, thus accelerating the training process significantly.

\vspace{-2mm}
\subsection{Enabling Seasonal Adaptivity in Policy via an Expertise-Based Mechanism}\label{ability1}
\begin{figure}[tb]
	\centering
		\includegraphics[width=\columnwidth]{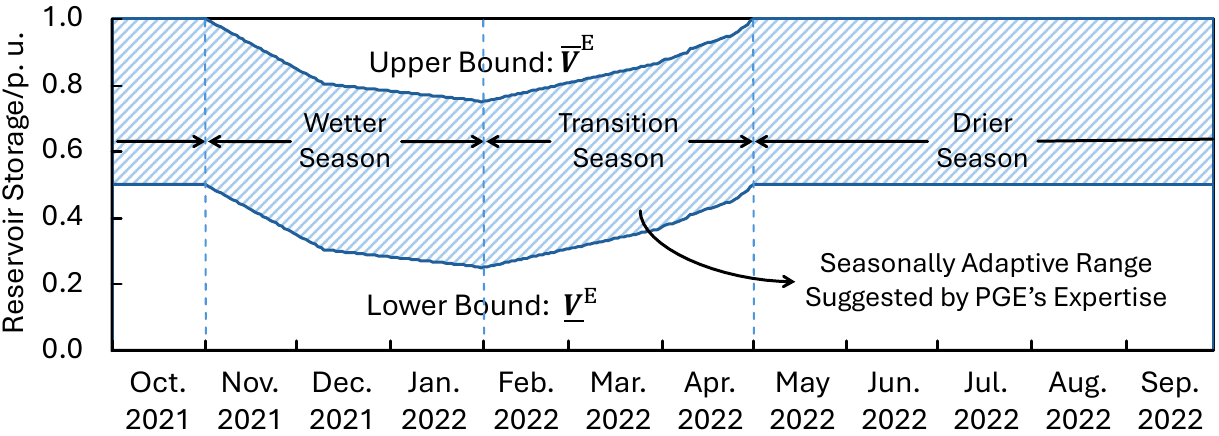}
  \vspace{-7mm}
	\caption{PGE's seasonality expertise over the 2022 water year.}\label{fig07}
\vspace{-6mm}
\end{figure}

Generally, S-CHPs have extensive expertise in achieving seasonal adaptivity. Taking PGE as an example, its operators ensure seasonal adaptivity by maintaining reservoir storage levels within a preset range shown in Fig.~\ref{fig07}. During the wetter season (November to January), reservoir storage remains appropriately low; in the subsequent transition season (February to April), reservoirs gradually refill to ensure sufficient storage by May in preparation for the upcoming drier season (May to October); throughout the drier season, the reservoirs maintain an adequate storage level.

In the game Space Impact \cite{SpaceImpact}, players navigate a spaceship flying through an uneven tunnel, accumulating rewards while avoiding crashes into the ceiling and floor. Inspired by the similarity between this game mechanism and the seasonal adaptivity requirements, a mechanism based on the expertise range shown in Fig.~\ref{fig07} is incorporated into the policy training.

Specifically, by leveraging the idea of game Space Impact, we use \eqref{NewState} and \eqref{logical} to enable seasonal adaptivity of the policy in a simple yet effective way.
Equation \eqref{NewState} augments the state $\boldsymbol{s}_{k}$ with additional $\bar{V}^{\text{E}}_{k}$ and $\underline{V}^{\text{E}}_{k}$, representing the expertise-based upper and lower bounds for day $k$. With this, the logical statement \eqref{logical} is deployed at each step after revealing $\tilde{\boldsymbol{V}}_{k}^{\smash{\raisebox{-0.5ex}{\scriptsize end}}}$, encouraging the policy to maintain  $\tilde{\boldsymbol{V}}_{k}^{\smash{\raisebox{-0.5ex}{\scriptsize end}}}$ within the range (i.e., avoid potential crashes). With this, if executing the planning strategy $\boldsymbol{a}_{k}$ causes $\tilde{\boldsymbol{V}}_{k}^{\smash{\raisebox{-0.5ex}{\scriptsize end}}}$ to fall outside $[\underline{V}^{\text{E}}_{k}, \bar{V}^{\text{E}}_{k}]$, the current episode terminates and the revenue for remaining days $k+1,...,K$ is set to zero, thus incurring a low annual profit for this episode. By feeding this profit information back to update parameters $\theta$, the policy $\pi_{\theta}$ gradually learns to generate planning strategies $\boldsymbol{a}_{k}$ that can successfully complete the entire water year (i.e., achieving seasonally adaptable $\tilde{\boldsymbol{V}}_{k}^{\smash{\raisebox{-0.5ex}{\scriptsize end}}}$) while receiving high profits in WEIM participation.
\begin{equation}\label{NewState}
\textstyle{\boldsymbol{s}_{k}=
[k, \hat{\boldsymbol{W}}_{k}, \hat{\boldsymbol{P}}_{k}^{\smash{\raisebox{-0.5ex}{\scriptsize vr}}}, \boldsymbol{\lambda}_{k}, \boldsymbol{V}^{\text{ini}}_{k}, \underline{V}^{\text{E}}_{k}, \bar{V}^{\text{E}}_{k}]}
\end{equation}
\begin{equation}\label{logical}
\begin{cases} 
\text{Continue episode}  & \displaystyle \text{if }  \underline{V}^{\text{E}}_{k} \leq \frac{\sum_{n=1}^{N}\tilde{V}^{\text{end}}_{n,k}}{\sum_{n=1}^{N}V^{\text{M}}_{n}} \leq \bar{V}^{\text{E}}_{k}, \\
\text{Terminate episode} & \text{otherwise}.
\end{cases}
\end{equation}

\vspace{-3mm}
\subsection{Training Acceleration via Multi-Parametric Programming}\label{ability2}
\vspace{-1mm}
Three properties of the rolling RT operation are noteworthy:
\begin{itemize}[noitemsep, topsep=0pt, parsep=0pt, partopsep=0pt, leftmargin=*, wide = 0pt]
\item
Individual RT operation runs only differ in the following right-hand-side parameters: generation plans $P_{t}^{\text{sch}\star}$, actual VRES availability $\tilde{P}^{\text{vr}}_{j_{t}}$, initial reservoir storage $\bar{\boldsymbol{V}}_{j_{t}}^{\smash{\raisebox{-0.5ex}{\scriptsize RT}}}$, scheduled inflows from upstream reservoirs $\bar{\boldsymbol{W}}_{j_{t}}^{\smash{\raisebox{-0.5ex}{\scriptsize RT,i}}}$, and natural inflows $\tilde{\boldsymbol{W}}_{j_{t}}$. They are collected in vector $\boldsymbol{\vartheta}$ \eqref{parameter}, with its space $\Theta$ defined in \eqref{ParameterSpace}. $P^{\text{vr,M}}$ is the installed VRES capacity, and $W_{n}^{\text{M}}$ is the maximum allowable inflow for reservoir $n$;
\begin{equation}\label{parameter}
\boldsymbol{\vartheta}=[P_{t}^{\text{sch}\star}, \tilde{P}^{\text{vr}}_{j_{t}}, \bar{\boldsymbol{V}}_{j_{t}}^{\smash{\raisebox{-0.5ex}{\scriptsize RT}}}, \bar{\boldsymbol{W}}_{j_{t}}^{\smash{\raisebox{-0.5ex}{\scriptsize RT,i}}}, \tilde{\boldsymbol{W}}_{j_{t}}]\,, \boldsymbol{\vartheta} \in \Theta;
\end{equation}
\begin{equation}\label{ParameterSpace}
\renewcommand{\arraystretch}{1.3}
\Theta=\left\{\mspace{-5mu}
\begin{array}{l}
\textstyle{0                \leq P_{t}^{\text{sch}\star}       \leq \sum\nolimits_{n \in \mathcal{N}} \sum\nolimits_{i \in \mathcal{I}_{n}} P^{\text{M}}_{ni} + P^{\text{vr,M}};}                 \\
\textstyle{0                \leq \tilde{P}^{\text{vr}}_{j_{t}}  \leq P^{\text{vr,M}};\,\,}\textstyle{V_{n}^{\text{m}} \leq \bar{V}_{nj_{t}}^{\text{RT}}    \leq V_{n}^{\text{M}},\,  \forall{n};}\\
\textstyle{0                \leq \bar{W}_{nj_{t}}^{\text{RT,i}},\,\,\tilde{W}_{nj_{t}} \leq W_{n}^{\text{M}},\,  \forall{n};}
\end{array}
\mspace{-10mu}
\right\}
\end{equation}

\item
The RT operation problem \eqref{RTModelA} is a MILP formulation that contains a moderate number of constraints, as each instance of \eqref{RTModelA} corresponds to a single SI;

\item
In \eqref{RTModelA}, only a subset of operational decision solutions needs to be retained for future use. Specifically, delivery deviation $P_{j_{t}}^{\text{RT,}\Delta}$ is needed to calculate the reward \eqref{Reward}. Meanwhile, water spillage $\boldsymbol{W}_{j_{t}}^{\smash{\raisebox{-0.05ex}{\scriptsize RT,ws}}}$, scheduled outflows  $\boldsymbol{W}_{j_{t}}^{\smash{\raisebox{-0.05ex}{\scriptsize RT,o}}}$, and end-of-SI storage volume $\boldsymbol{V}_{{j_{t}}+1}^{\smash{\raisebox{-0.05ex}{\scriptsize RT}}}$ serve as inputs to models \eqref{RTModelA} of future SIs. We use a vector $\boldsymbol{\xi}$ to represent them as in \eqref{varible}.
\begin{equation}\label{varible}
\boldsymbol{\xi}=[P_{j_{t}}^{\text{RT,}\Delta}, \boldsymbol{W}_{j_{t}}^{\text{RT,ws}}, \boldsymbol{W}_{j_{t}}^{\text{RT,o}}, \boldsymbol{V}_{{j_{t}}+1}^{\text{RT}}]
\end{equation}
\end{itemize}

These three properties motivate the development of an MPP-based approach to accelerate the DRL training process. It includes two major steps, as detailed below:

\subsubsection{Partitioning Parameter Space $\Theta$} We first combine the RT operation model \eqref{RTModelA} with the parameter space \eqref{ParameterSpace} to form a multi-parametric MILP (mp-MILP) problem \eqref{mpMILP}. Here, \eqref{mpMILP:1}-\eqref{mpMILP:3} stand for model \eqref{RTModelA}, where $\boldsymbol{c}$, $\boldsymbol{d}$, $\boldsymbol{A}$, $\boldsymbol{E}$, $\boldsymbol{b}$, and $\boldsymbol{F}$ are constant vectors and matrices; \eqref{mpMILP:4} represents \eqref{ParameterSpace}.
\begin{subequations}\label{mpMILP}
\begin{align}
&\textstyle{v^{\star}(\boldsymbol{\vartheta})= \min\nolimits_{\boldsymbol{x}^{\text{RT}}, \boldsymbol{y}^{\text{RT}}}}\,\, \boldsymbol{c}^{\top}\boldsymbol{x}^{\text{RT}} + \boldsymbol{d}^{\top}\boldsymbol{y}^{\text{RT}}                                                     \label{mpMILP:1} \\
&\text{s.t. } \boldsymbol{A}\boldsymbol{x}^{\text{RT}}\mspace{-3mu}+\mspace{-3mu}\boldsymbol{E}\boldsymbol{y}^{\text{RT}}\mspace{-3mu} \leq \mspace{-3mu}\boldsymbol{F} \boldsymbol{\vartheta}\mspace{-3mu}+\mspace{-4mu}\boldsymbol{b}; \boldsymbol{x}^{\text{RT}}\mspace{-5mu} \in \mspace{-5mu}\mathbb{R}^{p^{\text{RT}}}_{+}\mspace{-3mu},\mspace{-2mu}\, \mspace{-3mu}\boldsymbol{y}^{\text{RT}}\mspace{-5mu} \in \mspace{-5mu}\{0,\mspace{-2mu}1\}^{q^{\text{RT}}}; \label{mpMILP:3} \\
&\mspace{30mu}\boldsymbol{\vartheta} \in \Theta,\, \boldsymbol{\vartheta} \in \mathbb{R}^{2+3N}_{+};  \label{mpMILP:4}
\end{align}
\end{subequations}

\begin{figure}[tb]
	\centering
		\includegraphics[width=0.92\columnwidth]{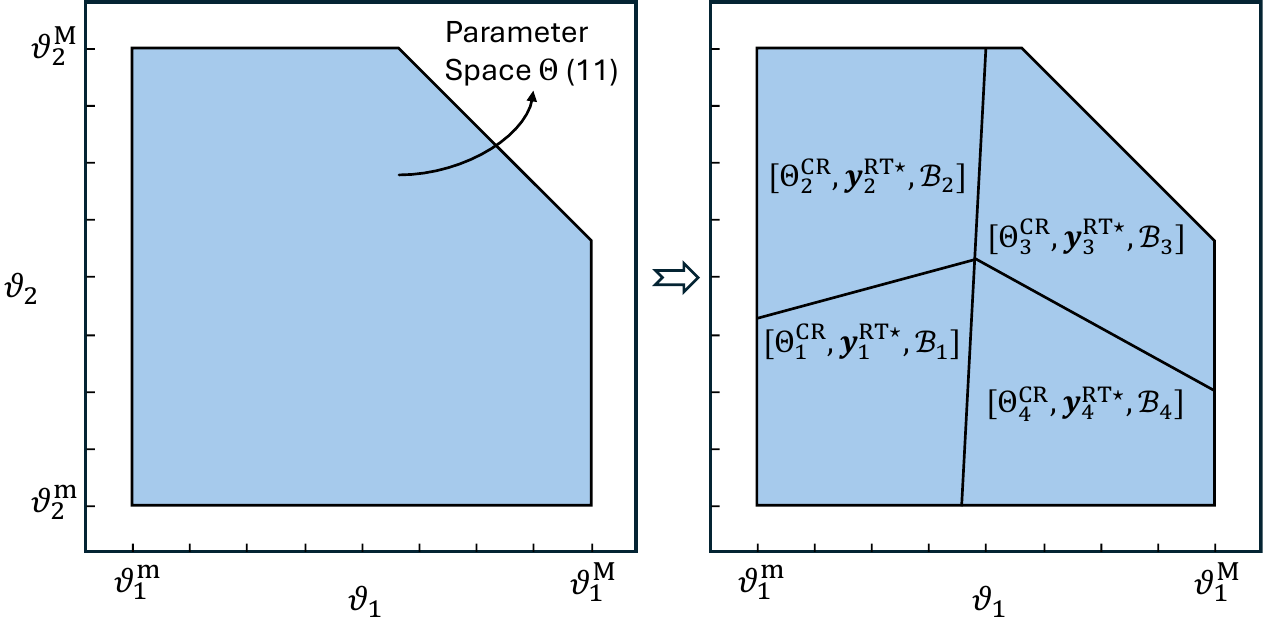}
  \vspace{-4mm}
	\caption{Illustrative example of partitioning a 2-dimensional parameter space.}\label{fig08}
 \vspace{-6mm}
\end{figure}

The iteration algorithm from reference \cite{Guo_mpMILP_app1} is used to solve the mp-MILP problem \eqref{mpMILP}, yielding a set of $R$ tuples $\{[\Theta_{1}^{\text{CR}}, \boldsymbol{y}^{\text{RT}\star}_{\text{1}}, \mathcal{B}_{1}],..., [\Theta_{R}^{\text{CR}}, \boldsymbol{y}^{\text{RT}\star}_{R}, \mathcal{B}_{R}]\}$. Each tuple consists of a critical region (CR) $\Theta_{r}^{\text{CR}}$, as well as an optimal binary solution $\boldsymbol{y}^{\text{RT}\star}_{r}$ and a set of binding inequality constraints (i.e., the active set \cite{mp_yufan2}) $\mathcal{B}_{r}$ that remain valid for $\forall \boldsymbol{\vartheta} \in \Theta_{r}^{\text{CR}}$. As shown in Fig.~\ref{fig08}, these CRs are polytopes and satisfy exhaustiveness ($\cup_{r=\text{1}}^{R} {\Theta}_{r}^{\text{CR}} = \Theta$) and disjointness (${\Theta}_{r}^{\text{CR}} \cap {\Theta}_{r^{\prime}}^{\text{CR}} = \emptyset $ for $r \neq r^{\prime}$). 

\subsubsection{Deriving Analytical Functions of Optimal Solutions} 
For each CR $r$, as the optimal binary solution $\boldsymbol{y}^{\text{RT}\star}_{r}$ remains valid for $\forall \boldsymbol{\vartheta} \in \Theta_{r}^{\text{CR}}$, solving $\boldsymbol{x}^{\text{RT}}$ in \eqref{mpMILP} for any $\boldsymbol{\vartheta} \in \Theta_{r}^{\text{CR}}$ can be equivalently represented as a linear programming (LP) model \eqref{step2} with the fixed $\boldsymbol{y}^{\text{RT}\star}_{r}$ \cite{mp_yufan3}.
Furthermore, as the active set $\mathcal{B}_r$ remains valid for $\forall \boldsymbol{\vartheta} \in \Theta_{r}^{\text{CR}}$, the optimal solution $\boldsymbol{x}^{\text{RT}\star}$ of \eqref{step2} must satisfy \eqref{ActiveSet}, where matrices $\boldsymbol{A}^{\text{AS}}_{r}$ and $\boldsymbol{F}^{\text{AS}}_{r}$ and vector $\boldsymbol{b}^{\text{AS}}_{r}$ collect the rows 
of $\boldsymbol{A}$, $\boldsymbol{F}$, and $(\boldsymbol{b}-\boldsymbol{E}\boldsymbol{y}^{\text{RT}\star}_{r})$
corresponding to the binding inequality constraints in $\mathcal{B}_r$.
\begin{subequations}\label{step2}
\begin{align}
&\textstyle{\min\nolimits_{\boldsymbol{x}^{\text{RT}}}\boldsymbol{c}^{\top}\boldsymbol{x}^{\text{RT}} + \boldsymbol{d}^{\top} \boldsymbol{y}^{\text{RT}\star}_{r}} \label{step2:A}\\
&\text{s.t. }\boldsymbol{A}\boldsymbol{x}^{\text{RT}} \leq \boldsymbol{F} \boldsymbol{\vartheta} + \boldsymbol{b} -\boldsymbol{E}\boldsymbol{y}^{\text{RT}\star}_{r};\, \boldsymbol{x}^{\text{RT}} \in \mathbb{R}^{p^{\text{RT}}}_{+}; \label{step2:B}\\
&\mspace{30mu}\boldsymbol{\vartheta} \in \Theta^{\text{CR}}_{r},\, \boldsymbol{\vartheta} \in \mathbb{R}^{2+3N}_{+}; \label{step2:C}
\end{align}
\end{subequations}

  \vspace{-3mm}
\begin{equation}\label{ActiveSet}
\boldsymbol{A}^{\text{AS}}_{r}\boldsymbol{x}^{\text{RT}\star} = \boldsymbol{F}^{\text{AS}}_{r}\boldsymbol{\vartheta}+\boldsymbol{b}^{\text{AS}}_{r}
\end{equation}

If the LP model \eqref{step2} satisfies both linear independence constraint qualification (LICQ) and strict complementary slackness (SCS) conditions (i.e., model \eqref{step2} is non-degenerate),
matrix $\boldsymbol{A}^{\text{AS}}_{r}$ is a full rank square matrix according to \cite{book}.
With this, the optimal solution $\boldsymbol{x}^{\text{RT}\star}$ of \eqref{step2} can be expressed as a linear function of parameter $\boldsymbol{\vartheta}$ \cite{Guo_mpMILP_app1}, as shown in \eqref{BacicSens}.
If the LP model \eqref{step2} fails to simultaneously satisfy LICQ and SCS, $\boldsymbol{A}^{\text{AS}}_{r}$ becomes a non-square matrix and is not directly invertible. In such cases, the methods in \cite{inverseTrick, redundant} can be applied to find an approximation for ${\boldsymbol{A}^{\text{AS}}_{r}}^{-1}$, and consequently \eqref{BacicSens} delivers an approximation of the optimal solution $\boldsymbol{x}^{\text{RT}\star}$.

  \vspace{-3mm}
\begin{equation}\label{BacicSens}
\boldsymbol{x}^{\text{RT}\star}={\boldsymbol{A}^{\text{AS}}_{r}}^{-1}\boldsymbol{F}^{\text{AS}}_{r}\boldsymbol{\vartheta}
+{\boldsymbol{A}^{\text{AS}}_{r}}^{-1}\boldsymbol{b}_{r}^{\text{AS}}
\end{equation}

Next, by identifying the elements of $\boldsymbol{x}^{\text{RT}\star}$ in \eqref{BacicSens} that form $\boldsymbol{\xi}$ and collecting the corresponding rows of ${\boldsymbol{A}^{\text{AS}}_{r}}^{-1}\boldsymbol{F}^{\text{AS}}_{r}$ and ${\boldsymbol{A}^{\text{AS}}_{r}}^{-1}\boldsymbol{b}_{r}^{\text{AS}}$ into $\boldsymbol{A}^{\prime}_r$ and $\boldsymbol{b}^{\prime}_r$, the optimal solution $\boldsymbol{\xi}^{\star}$ in CR $\Theta_{r}^{\text{CR}}$ can be expressed as a linear function \eqref{OneIfthen}.
\begin{equation}\label{OneIfthen}
\boldsymbol{\xi}^{\star}=\boldsymbol{A}^{\prime}_{r}\boldsymbol{\vartheta} + \boldsymbol{b}^{\prime}_{r}
\end{equation}

Finally, by applying the above process to all $R$ tuples, a conditional linear function \eqref{Ifthen} can be generated to express $\boldsymbol{\xi}^{\star}$ over the entire parameter space $\Theta$.
\begin{equation}\label{Ifthen}
\left\{ \begin{array}{ll}
\boldsymbol{\xi}^{\star}=\boldsymbol{A}^{\prime}_{1}\boldsymbol{\vartheta} + \boldsymbol{b}^{\prime}_{1}&\text{if } \boldsymbol{\vartheta} \in {\Theta}_{1}^{\text{CR}};\\
                          \vdots                            &\\
\boldsymbol{\xi}^{\star}=\boldsymbol{A}^{\prime}_{R}\boldsymbol{\vartheta} + \boldsymbol{b}^{\prime}_{R}&\text{if } \boldsymbol{\vartheta} \in {\Theta}_{R}^{\text{CR}};\\
\end{array}\right.
\end{equation}

The analytical expression \eqref{Ifthen} allows the training process to quickly unveil $\boldsymbol{\xi}^{\star}$ without
 explicitly solving millions of MILP-based RT operation problems \eqref{RTModelA}: given a parameter vector $\boldsymbol{\vartheta}$, check the conditions in \eqref{Ifthen} to identify its CR and perform the corresponding linear function calculations to unveil $\boldsymbol{\xi}^{\star}$. As a result, the training process is significantly accelerated.

\section{Case Studies on the PGE System}\label{Sec4}
\subsection{The VS-CHP of PGE and Experimental Settings}
\begin{figure}[b]
	\centering
    \vspace{-4mm}
		\includegraphics[width=\columnwidth]{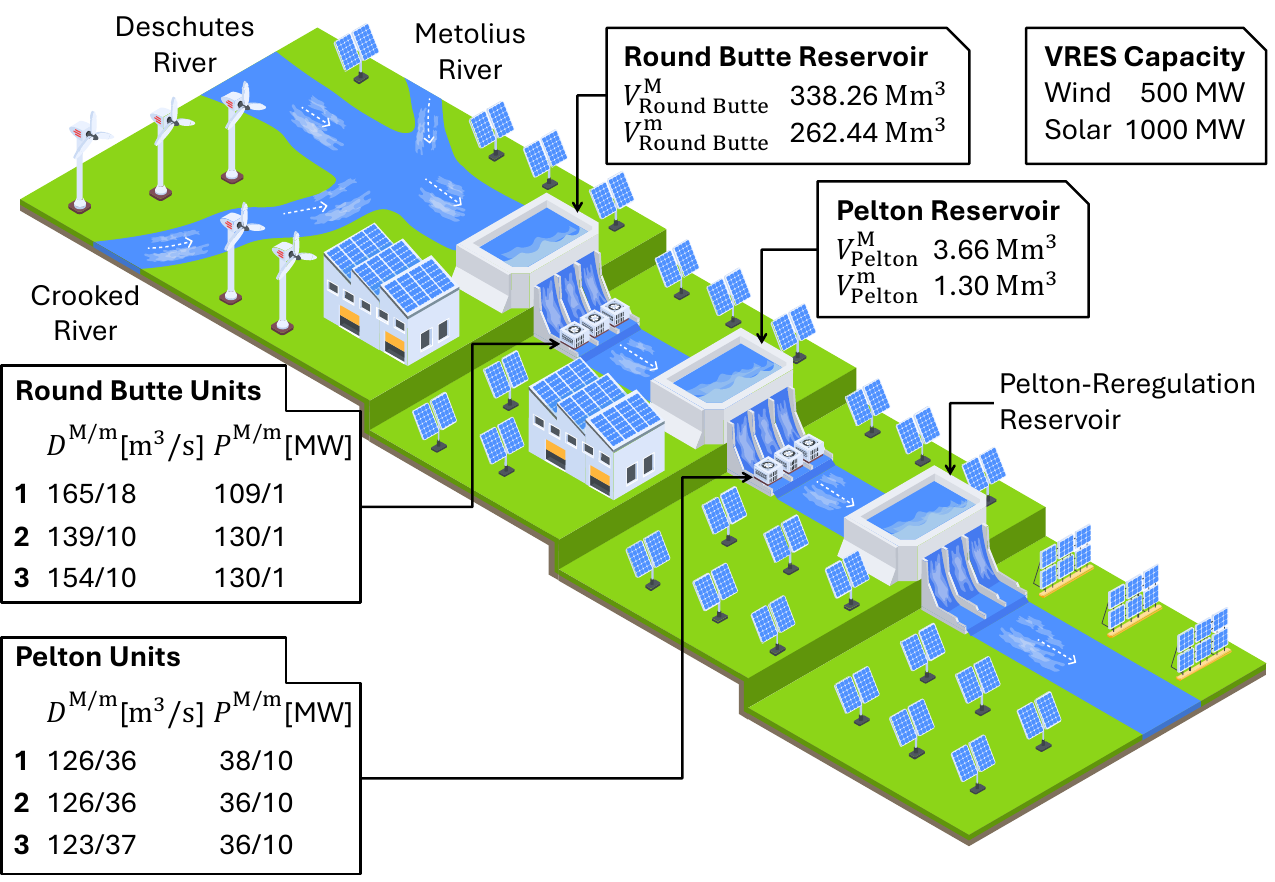}
        \vspace{-6mm}
	\caption{Illustration of PGE's VS-CHP.}\label{fig09}
\end{figure}

This section uses one of PGE's VS-CHPs to evaluate the presented framework. As shown in Fig.~\ref{fig09}, the selected VS-CHP is based on the Pelton-Round Butte project in Central Oregon, which has a hydropower capacity of 479 MW and integrates a VRES capacity of 1,500 MW. The Pelton-Round Butte project comprises three cascaded reservoirs: the upstream Round Butte (RB), the middle Pelton (PT), and the downstream PT-Reregulation. The RB and PT reservoirs are each equipped with three hydro units, while the PT-Reregulation reservoir serves to stabilize the downstream flow instead of generating electricity. The system receives natural inflows through RB, primarily from three upstream branches: Crooked River, Deschutes River, and Metolius River.

A PGE dataset spanning the 2022 and 2023 water years is utilized. The data from the 2022 water year (2021-10-01 to 2022-09-30) are used for training, and the data from the 2023 water year (2022-10-01 to 2023-09-30) are used for testing. Therefore, $K$ is 365. The penalty coefficient $C^{\Delta}_{j_{t}}$ for under-generation is set to 10 times RT LMP of SI $j_{t}$.

In recognizing its sample efficiency and stochastic nature, the SAC algorithm \cite{QT_SAC_1} is selected as the DRL algorithm within the presented framework. Thus, the framework has two specific instances named SAC-ES and SAC-WV, respectively, using end-of-day storage and water values as medium-term planning strategies in policies $\pi_{\theta}$. All experiments are conducted on a PC with Intel Core i9-11900K processor@3.50 GHz. LP and MILP problems are solved via Gurobi 10, called by Pyomo. The SAC algorithm is implemented in PyTorch \cite{QT_SAC_2}, and the environment is developed using Gymnasium. To focus on evaluating the effectiveness of medium-term planning strategies from the presented framework, only a few hyperparameters listed in Table~\ref{Tab01} are customized, while the remaining ones keep the default settings in OpenAI Spinning Up \cite{openai_sac}. The source code can be accessed at \cite{DRLCode} for reproducibility.

\begin{table}[tb]
\renewcommand{\arraystretch}{0.9}
\footnotesize
    \centering
    \caption{Customized Hyperparameters}\label{Tab01}
    \vspace{-2mm}
    \begin{tabular}{cc}
\toprule
Hyperparameter Name            &Setting   \\
\midrule
Number of Training Steps       &182,500 (500$\times$365)\\
Size of Replay Buffer          &365,000 (2$\times$182500)\\
Updating Frequency             &Update after each episode\\
\bottomrule
    \end{tabular}
    \vspace{-4mm}
\end{table}

\vspace{-2mm}
\subsection{Training Performance}
\vspace{-1mm}

In our experiments, both SAC-ES and SAC-WV take about 40 hours to complete the 182,500 training steps. In the final episodes, both SAC-ES and SAC-WV can successfully complete the 2022 water year, meaning that they ultimately learn to achieve seasonal adaptivity throughout this water year without crashing outside the range shown in Fig.~\ref{fig07}.

\begin{figure}[b]
	\centering
    \vspace{-6mm}
		\includegraphics[width=0.95\columnwidth]{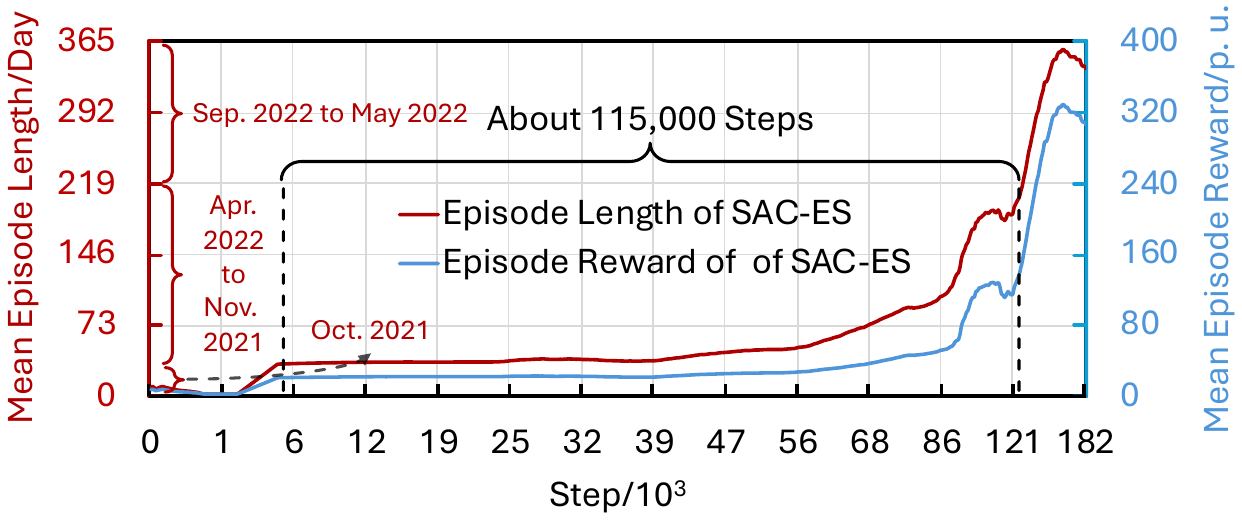}
   \vspace{-4mm}
 \caption{The episode performance of SAC-ES.}\label{fig10}
\end{figure}

The training curves of SAC-ES and SAC-WV also confirm that they can finally adapt to seasonality. Fig.~\ref{fig10} illustrates this point using SAC-ES as an example. SAC-ES requires about 115,000 steps to learn to pass through the wetter and transition seasons (November 2021 to April 2022), which is significantly higher than that for the drier season (October 2021 and May-November 2022). This is because the seasonally adaptive range for the wetter and transition seasons presents a V-shape (see Fig.~\ref{fig07}), while the drier season remains a box shape. As a result, the policy needs more iterations to learn to traverse the V-shaped ``tunnel."

The benefit of the proposed MPP-based acceleration method is highlighted by comparing average per-step training times without and with the acceleration method,
which are 69.2 seconds and 0.8 seconds with a 98.8\% reduction. This result indicates that constructing a computationally efficient environment is essential for leveraging off-the-shelf DRL algorithm libraries to solve medium-term VS-CHP planning and similar tasks. Without acceleration methods, training policies for tasks that involve numerous optimization problems in millions of steps would be prohibitively time-consuming.

\vspace{-2mm}
\subsection{Comparison of Different Medium-Term Planning Methods}
\vspace{-1mm}
The method named PGE-P is used as the benchmark to compare the effects of different medium-term planning methods. PGE-P represents the experience-based approach currently used by PGE to determine medium-term planning strategies for its Pelton-Round Butte S-CHP, which sets $V_{n}^{\text{ini}} = \hat{V}_{n}^{\text{end}}$.

\setlength{\tabcolsep}{3.5pt}
\begin{table}[tb]
\renewcommand{\arraystretch}{0.85}
	\caption{Comparison of revenues in 2023 Water Year /\$$10^{6}$}\label{tab02}
     \vspace{-2mm}
	\centering
	\footnotesize
\begin{tabular}{cccccc}
\toprule
\multirow{1}{*}{\shortstack[c]{Period}}&\multirow{1}{*}{Method}&\multirow{1}{*}{\shortstack[c]{Imbalance Charge}} &\multirow{1}{*}{\shortstack[c]{Gross Revenue}}&\multirow{1}{*}{\shortstack[c]{Net Revenue}}\\
\midrule
\multirow{3}{*}{\shortstack[c]{Entire \\Water Year}} & SAC-ES    & 15.2        & 341.2           & 326.0             \\
                                                   & SAC-WV    & 31.0        & 340.7           & 309.7             \\
                                                   & PGE-P     & 25.5        & 348.8           & 323.3              \\
\midrule
\multirow{3}{*}{\shortstack[c]{Drier \\Season}}      & SAC-ES    & 9.5         & 146.7           & 137.2               \\
                                                   & SAC-WV    & 10.8        & 145.1           & 134.3               \\
                                                   & PGE-P     & 19.1        & 153.2           & 134.1               \\
\midrule
\multirow{3}{*}{\shortstack[c]{Wetter \\Season}}     & SAC-ES    & 3.0         & 127.7           & 124.7               \\
                                                   & SAC-WV    & 17.9        & 127.9           & 110.0               \\
                                                   & PGE-P     & 4.2         & 128.8           & 124.6               \\
\midrule
\multirow{3}{*}{\shortstack[c]{Transition \\Season}} & SAC-ES    & 2.7         & 66.8            & 64.1                \\
                                                   & SAC-WV    & 2.3         & 67.7            & 65.4                \\
                                                   & PGE-P     & 2.2         & 66.8            & 64.6                \\
\bottomrule
\end{tabular}
   \vspace{-5mm}
\end{table}

Table~\ref{tab02} compares the monetary performance of SAC-ES, SAC-WV, and PGE-P over the 2023 water year. The second row shows that, in terms of annual net revenue, SAC-ES outperforms PGE-P by 0.8\%, while SAC-WV is 4.2\% lower than PGE-P. Moreover, both SAC-ES and SAC-WV cause lower gross annual revenue than PGE-P, implying that they prefer higher end-of-day storage levels than PGE-P. This is noteworthy because PGE-P, which is typically considered conservative for hydropower scheduling, appears relatively aggressive when applied to the VS-CHP. Indeed, our numerical results indicate that hydropower dispatches in the generation plans of SAC-ES are generally lower than those of PGE-P. That is, SAC-ES holds greater hydropower ramping capacities as flexible resources to compensate for VRES shortages during the RT stage, leading to much lower imbalance charges.

Additionally, the third to fifth rows of Table~\ref{tab02} present the monetary performance in the three water seasons. The results show that, although the financial advantages of SAC-ES and SAC-WV are not significant in wetter and transition seasons, they both achieve higher net revenue than PGE-P in the drier season.  Recall Fig.~\ref{fig10} that SAC-ES spends only one-third of the 182,500 training steps to complete the drier season learning that spans half of the 2022 water year. These in-sample and out-of-sample results suggest that SAC-ES, with end-of-day storage as the planning strategy, is particularly effective at improving net revenue for VS-CHPs in the drier season.

\begin{figure}[tb]
	\centering
		\includegraphics[width=\columnwidth]{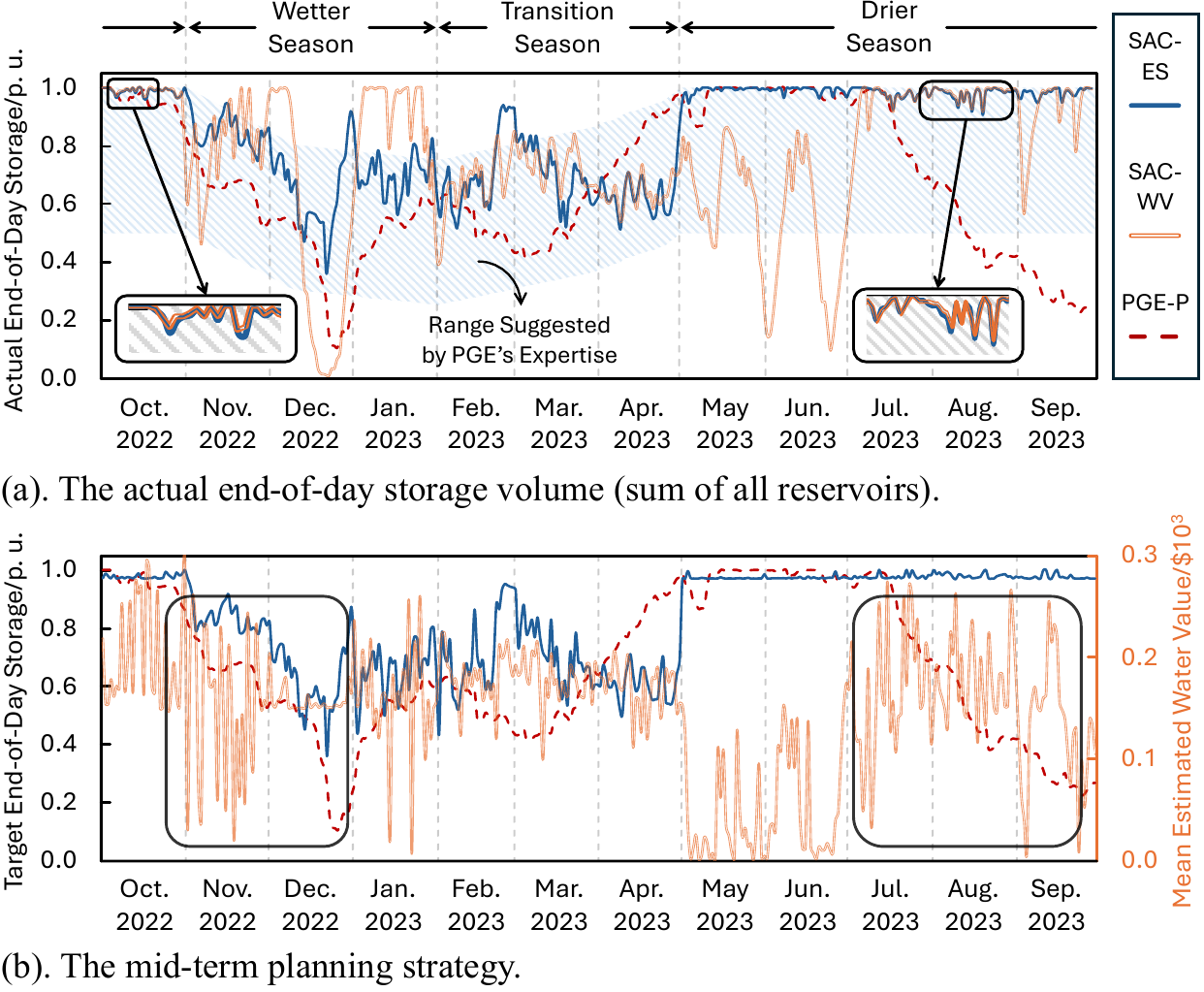}
     \vspace{-7mm}
	\caption{Comparison of SAC-ES, SAC-WV, and PGE-P in 2023 water year.}\label{fig11}
   \vspace{-6mm}
\end{figure}

Fig.~\ref{fig11} further compares the end-of-day storage details. Fig.~\ref{fig11}(a) shows that only SAC-ES can maintain the actual end-of-day storage above the lower bound of the seasonally adaptive range throughout the year. As compared, SAC-WV falls below this lower bound three times but recovers within two weeks each time. PGE-P experiences two significant declines: the first decline starts in late October 2022, breaks through the lower bound in December 2022, and then bounces back and recovers to above the lower bound in January 2023; the second decline starts in mid-July 2023 and fails to recover by the end of the water year. PGE-P's two declines are caused by over-predictions of VRES availability, which occur frequently during the two decline periods. These deviations require the VS-CHP to use more water than expected to meet its generation plans accepted by WEIM, resulting in actual storage levels being significantly lower than the planned targets. Recall that PGE-P sets $V_{n}^{\text{ini}} = \hat{V}_{n}^{\text{end}}$. If VRES availability is frequently over-predicted, both the target and actual end-of-day storage of PGE-P will gradually decrease, as shown in Fig.~\ref{fig11}(b). Thus, PGE-P may fail to meet the seasonal adaptivity requirements. In comparison, the planning strategy (i.e., target end-of-day storage) of SAC-ES exhibits seasonality consistent with the suggested range: maintaining high storage in the drier season, gradually declining in the wetter season, and increasing in the transition season. More importantly, Fig.~\ref{fig11}(a) shows that SAC-ES can quickly return to the range after exceeding the upper bound, implying its sensitivity to this out-of-range information. Considering that falling below lower bounds induces more severe monetary consequences than exceeding upper bounds, these observations clearly show incorporating the expertise-based mechanism enables SAC-ES to achieve better seasonal adaptability than PGE-P.

It is also noteworthy that the actual end-of-day storage volumes of SAC-ES and SAC-WV overlap several times, such as the two zoomed-in periods in Fig.~\ref{fig11}(a). This implies that, given the same state $\boldsymbol{s}_k$, SAC-ES and SAC-WV would aim to achieve similar actual end-of-day storage levels; nevertheless, SAC-WV sometimes deviates significantly from its target, resulting in more fluctuations in its actual end-of-day storage profile. Furthermore, Fig.~\ref{fig11}(b) shows that the planning strategies provided by SAC-WV (i.e., the water values) exhibit more frequent fluctuations than the planning strategies of SAC-ES. These differences arise because SAC-WV influences actual end-of-day storage by adjusting the objective coefficient of the self-scheduling model \eqref{SModel}, whereas SAC-ES does so via hard constraint \eqref{HModel:4}. Ultimately, by directly affecting storage more straightforwardly, SAC-ES exhibits greater stability and better performance than SAC-WV.

   \vspace{-3mm}
\subsection{Effect of Hydro-VRES Ratio on Method Performance}
The capacities of hydropower and VRES in PGE's system are 479 MW and 1,500 MW, resulting in a hydro-VRES ratio of 1:3.13. To analyze the impact of the hydro-VRES ratio on the performance of SAC-ES, this subsection reduces the VRES capacity to 1,125 MW and 750 MW, corresponding to hydro-VRES ratios of 1:2.35 and 1:1.57, respectively.

\begin{table}[tb]
\renewcommand{\arraystretch}{0.85}
	\caption{Effect of Hydro-VRES Ratio on Revenue in Water Year 2023}\label{tab03}
     \vspace{-2mm}
	\centering
	\footnotesize
\begin{tabular}{cccc}
\toprule
\multirow{2.7}{*}{\shortstack[c]{Hydro-VRES Ratio}} & \multicolumn{2}{c}{Annual Net Revenue/$\$10^{6}$}            &\multirow{2.7}{*}{\shortstack[c]{Improvement of\\SAC-ES over PGE-P}}  \\
\cmidrule{2-3}
             & SAC-ES                      & PGE-P                          & \\
\midrule
1:3.13      & \text{325.8}  & \text{323.3} & +0.84\%\\
\midrule
1:2.35      & \text{276.0}  & \text{278.8} & -1.02\% \\
\midrule
1:1.57      & \text{227.0}  & \text{231.4} & -1.94\% \\
\bottomrule
\end{tabular}
     \vspace{-3mm}
\end{table}

\begin{figure}[tb]
	\centering
		\includegraphics[width=\columnwidth]{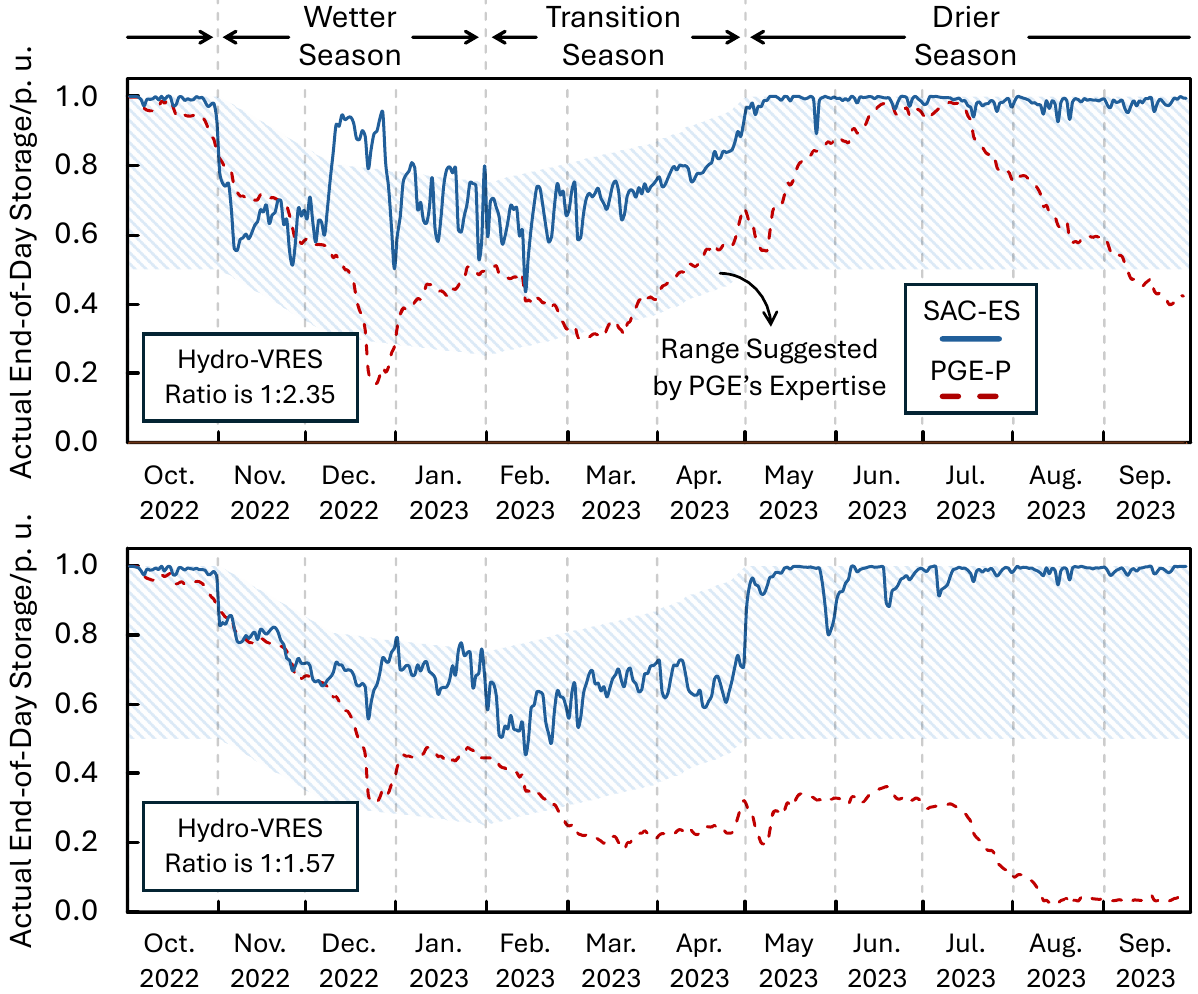}
     \vspace{-6mm}
	\caption{Effect of hydro-VRES ratios on actual end-of-day storage.}\label{fig12}
   \vspace{-3mm}
\end{figure}

Table~\ref{tab03} compares net revenues under different hydro-VRES ratios, revealing that the edge of SAC-ES diminishes and can even become negative as the hydro-VRES ratio decreases. Fig.~\ref{fig11}(a) and  Fig.~\ref{fig12} further show the effect of hydro-VRES ratios on the actual end-of-day storage, illustrating that the end-of-day storage profiles of PGE-P drop as the ratio decreases. 
This explains the negative improvement seen in Table~\ref{tab03}\textemdash a lower hydro-VRES ratio requires more frequent use of hydropower to smooth out VRES fluctuations during RT operations, ultimately making PGE-P discharges more water and thus leading to higher revenues than SAC-ES. Indeed, this shows SAC-ES possesses the same challenge in common machine learning (ML) applications, i.e., ML does not always yield benefits. Nevertheless, this issue could be partially mitigated through well-tuned hyperparameters.

In summary, PGE-P can achieve higher revenues under lower hydro-VRES ratios, but this comes at the price of violating seasonal adaptivity requirements. In comparison, SAC-ES slightly compromises the revenues (by approximately 1\%) but consistently ensures the end-of-storage requirements suggested by PGE's expertise, regardless of hydro-VRES ratios.

   \vspace{-3mm}
\subsection{Influence of Contextual Factors on Planning Strategies}
To further understand the role of contextual information \eqref{NewState} in the DRL-based decision-making for medium-term planning, this subsection adopts Shapley additive explanations (SHAP) values \cite{SHAP} to quantitatively measure how individual inputs affect the outputs of ML models. A larger SHAP value indicates a stronger positive/negative influence of a specific contextual factor on medium-term planning strategies. 

\begin{figure}[tb]
	\centering
		\includegraphics[width=\columnwidth]{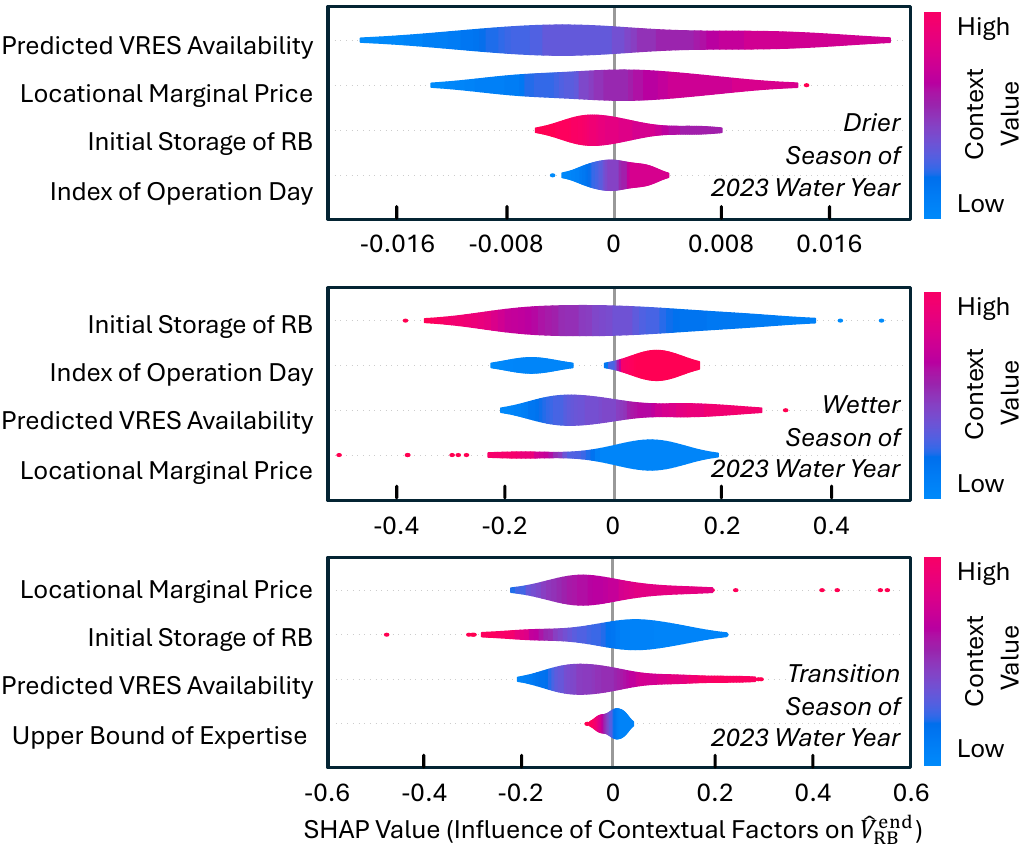}
     \vspace{-7mm}
	\caption{The four most influential context factors for $\hat{V}^{\text{end}}_{\text{RB}}$ of SAC-ES.}\label{fig13}
   \vspace{-6mm}
\end{figure}

Fig.~\ref{fig13} shows the four most influential short-term contextual factors that affect the target end-of-day storage of RB (i.e., $\hat{V}^{\text{end}}_{\text{RB}}$) yielded by SAC-ES's policy in each of the drier, wetter, and transition seasons of the 2023 water year. These factors are ranked based on their overall contribution to $\hat{V}^{\text{end}}_{\text{RB}}$ over the 2023 water year. The following observations are noteworthy:

\noindent \textit{\textbf{i).}}
The overall contribution of individual contextual factors varies in the three water seasons. However, three factors\textemdash Predicted VRES Availability, LMP, and Initial Storage of RB\textemdash consistently appear in the top four influential factors;

\noindent \textit{\textbf{ii).}}
The x-axis shows that SHAP values in the drier season have a smaller order of magnitude than the other two seasons. This implies that SAC-ES places less emphasis on the context during drier seasons as it mainly focuses on maintaining sufficient end-of-day storage levels, as shown in Fig.~\ref{fig11}(b);

\noindent \textit{\textbf{iii).}}
The rows of Predicted VRES Availability in all three seasons exhibit a similar trend\textemdash a higher predicted VRES power leads to a higher end-of-day storage target. This indicates that when the VRES is predicted to be abundant, less hydropower is allocated for $\boldsymbol{P}^{\text{sch}\star}$, preserving more flexible hydropower resources to offset VRES fluctuations during RT operations; 

\noindent \textit{\textbf{iv).}}
The rows of Initial Storage of RB show that a higher $V_{\text{RB}}^{\text{ini}}$ causes a lower $\hat{V}^{\text{end}}_{\text{RB}}$. In comparison, PGE-P consistently sets $V_{\text{RB}}^{\text{ini}}=\hat{V}^{\text{end}}_{\text{RB}}$. This is why SAC-ES can effectively ensure seasonal adaptivity for VS-CHP, but PGE-P fails to do so. Additionally, since $V_{\text{RB}}^{\text{ini}}$ significantly influences $\hat{V}^{\text{end}}_{\text{RB}}$, SAC-ES's decision-making can be viewed as a contextual customization of PGE-P strategies tailored for VS-CHP operations;

\noindent \textit{\textbf{v).}}
The influences of LMPs in the three water seasons are not in the same direction. This suggests that, while seeking to maximize net revenue \eqref{Reward}, SAC-ES also emphasizes seasonal adaptivity as a more critical consideration. This observation confirms the effectiveness of the expertise-based mechanism.

In summary, the above five points clearly demonstrate that short-term contextual information plays a vital role in the medium-term planning of SAC-ES. For instance, when aware that VRES power is predicted to be abundant and the beginning-of-day storage level is low, SAC-ES can set proper target end-of-day storage levels to ensure that the actual end-of-day storage level remains above the expertise-recommended lower bound while preserving enough flexible hydropower resources to manage VRES fluctuations during RT operations.

\section{Conclusions}\label{Sec5}
This paper presents a DRL-based medium-term planning framework for VS-CHPs. The framework effectively leverages short-term operation context to inform medium-term planning strategies and evaluates medium-term planning strategies based on their effectiveness in guiding multi-step short-term operations in wholesale market participation with enhanced revenues. 
To meet the practical needs, the framework includes an expertise-based mechanism in the DRL algorithm to achieve seasonal adaptivity of the planning strategy, as well as an MPP-based method to accelerate the DRL training. Numerical case studies on PGE's VS-CHP lead to the following conclusions:
\begin{itemize}[noitemsep, topsep=0pt, parsep=0pt, partopsep=0pt, leftmargin=*, wide = 0pt]


\item
The presented framework can achieve 0.8\% higher annual net revenue with better seasonally adaptive storage levels than PGE's current practice, and end-of-day storage is a more effective form of medium-term planning strategies than water values in the presented framework;

\item
The MPP-based acceleration method can reduce DRL training time by 98.8\%, making the training process computationally affordable;

\item
The contextual information from short-term operations, especially predicted VRES availability and initial storage levels, is crucial to enable the presented framework to generate appropriate medium-term planning strategies.

\end{itemize}

Future work could incorporate risk-averse considerations into the framework so as to handle extreme weather conditions.

\appendix
Although the in-house decision-making process of different VS-CHPs could vary, their underlying principles are similar to the operation models described here.

\subsection{Self-Scheduling Models}\label{DAModel}
We use stochastic self-scheduling models that incorporate up-to-date information (e.g., LMPs and VRES predictions) and medium-term planning strategies to determine $\boldsymbol{P}^{\text{sch}}$. A set $\mathcal{W}$, consisting of $|\mathcal{W}|$ stochastic VRES scenarios with a probability $p_w$ of $1/|\mathcal{W}|$ each, is used in the models. The medium-term planning strategy can take the form of end-of-day storage targets to act as hard boundaries or water values to provide soft guidance on water usage in the model.

\subsubsection{Self-Scheduling Model With Target End-of-Day Storage Volumes}\label{DAModel1} The self-scheduling model with the end-of-day storage target $\hat{V}_{n}^{\text{end}}$ as the medium-term planning strategies is shown in \eqref{HModel}. The objective function \eqref{HModel:1} maximizes the gross revenue minus the expected under-generation penalty. Regarding constraints, \eqref{HModel:2} defines the gross revenue; \eqref{HModel:3} limits generation plans; \eqref{HModel:4} enforces the target end-of-day storage volume following the medium-term planning strategy; \eqref{HModel:5} specifies the initial storage; \eqref{HModel:6} calculates the penalty for under-generation; \eqref{HModel:7} indicates that the generated power is from hydropower and VRES; \eqref{HModel:8} limits VRES power by its predicted availability; \eqref{HModel:9} restricts storage within allowed boundaries; \eqref{HModel:10} describes storage evolution considering water flow time delay $\delta_{mn}$ between consecutive reservoirs $m$ and $n$; \eqref{HModel:11} defines the relationship between discharge rate and hydropower generation via $\mathcal{F}^{\text{DtP}}(\cdot)$, which is modeled via a piece-wise linearized function \cite{Arild_TSTE_SDDP_Aggregation_EODS, Arild_HHVER_Detailed_Long_Term_IET} to convert discharge rate (in $\text{m}^{\text{3}}$/s) into power (in MW); \eqref{HModel:12} and \eqref{HModel:13} limit the discharge rate and hydropower generation level.
\begin{subequations}\label{HModel}
\begin{flalign}
&\textstyle{\max\nolimits_{\Phi}}
\textstyle{\sum\nolimits_{t \in \mathcal{T}}G_{t} - p_{w}\sum\nolimits_{w \in \mathcal{W}}\sum\nolimits_{t\in \mathcal{T}} C_{wt}}                                     \mspace{-180mu}&\notag\\
&\text{where } \Phi=\{
\boldsymbol{C}, \boldsymbol{D}, \boldsymbol{G}, \boldsymbol{I}, \boldsymbol{P}, \boldsymbol{P}^{\text{act/cmit/vr}},
\boldsymbol{V}, \boldsymbol{W}^{\text{ws/i/o}}\}                  \mspace{-180mu}&\label{HModel:1}\\
&\textstyle{\text{s.t. }G_{t}=\sigma^{\text{H}} \lambda_{t}P_{t}^{\text{sch}};}
                                                                   \mspace{-180mu}& \label{HModel:2}\forall       t;\\
&\textstyle{\mspace{9mu} 0 \leq P^{\text{sch}}_{t} \leq \sum\nolimits_{n \in \mathcal{N}} \sum\nolimits_{i \in \mathcal{I}_{n}} P^{\text{M}}_{ni} + \hat{P}^{\text{vr}}_{t},}
                                                                    \mspace{-180mu}& \label{HModel:3} \forall       t;\\
&\textstyle{\mspace{9mu}V_{w, n,T+1}=\hat{V}_{n}^{\text{end}},}    \mspace{-180mu}& \label{HModel:4}\forall w,n    ;\\
&\textstyle{\mspace{9mu}V_{w,n,1}=V_{n}^{\text{ini}},}              \mspace{-180mu}& \label{HModel:5}\forall w,n    ;\\
&\textstyle{\mspace{9mu}C_{wt} = C^{\text{ud}}(P_{t}^{\text{sch}} - P_{wt}^{\text{act}}),\,P^{\text{sch}}_{t} \geq P_{wt}^{\text{act}},}                                                       \mspace{-180mu}& \label{HModel:6}\forall w,t    ;\\
&\textstyle{\mspace{9mu}P^{\text{act}}_{wt} = \sum\nolimits_{n \in \mathcal{N}} \sum\nolimits_{i \in \mathcal{I}_{n}}P_{wnit} + P^{\text{vr}}_{wt},}                                     \mspace{-180mu}& \label{HModel:7}\forall w,    t;\\
&\textstyle{\mspace{9mu}0 \leq P^{\text{vr}}_{wt} \leq \hat{P}^{\text{sce,vr}}_{wt},}
                                                                    \mspace{-180mu}& \label{HModel:8}\forall w,    t;\\
&\textstyle{\mspace{9mu}V^{\text{m}}_{n} \leq V_{wnt} \leq V^{\text{M}}_{n},}
                                                                    \mspace{-180mu}& \label{HModel:9}\forall w,n,  t;\\
&\textstyle{\mspace{9mu}V_{w,n,t+1}\mspace{-4mu}=\mspace{-4mu}V_{wnt}+ \hat{W}_{nt}+W^{\text{i}}_{wnt}-W^{\text{o}}_{wnt},}
                                                                    \mspace{-180mu}& \notag                           \\
&\textstyle{\mspace{9mu}W^{\text{i}}_{wnt}\mspace{-4mu}=\mspace{-4mu}\sum\nolimits_{m \in \bar{\mathcal{N}}_{n}}(\sum\nolimits_{i \in \mathcal{I}_{m}}\alpha D_{wmi(t-\delta_{mn})}+W^{\text{ws}}_{wm(t-\delta_{mn})}}),       \mspace{-180mu}& \notag\\
&\textstyle{\mspace{9mu}W^{\text{o}}_{wnt} \mspace{-4mu}= \mspace{-4mu}\sum\nolimits_{i \in \mathcal{I}_{n}}\mspace{-4mu} \alpha D_{wnit} + W^{\text{ws}}_{wnt},W^{\text{ws}}_{wnt} \geq 0,}                    \mspace{-180mu}& \label{HModel:10}\forall w,\mspace{-1mu}n,\mspace{-1mu}t;\\
&\textstyle{\mspace{9mu}P_{wnit} \mspace{-4mu}=\mspace{-4mu} \mathcal{F}^{\text{DtP}}\mspace{-2mu}(\mspace{-2mu} D_{wnit},\mspace{-2mu}I_{wnit}\mspace{-2mu} )\mspace{-2mu}, I_{wnit}\mspace{-3mu} \in \mspace{-3mu} \{0, \mspace{-1mu} 1\mspace{-1mu} \},} 
                                                                    \mspace{-180mu}& \label{HModel:11}\mspace{18mu}\forall w,\mspace{-1mu}n\mspace{-1mu},\mspace{-1mu}i\mspace{-1mu},\mspace{-1mu}t;\\
&\textstyle{\mspace{9mu}P^{\text{m}}_{ni} I_{wnit} \leq P_{wnit} \leq P^{\text{M}}_{ni} I_{wnit},}
                                                                    \mspace{-180mu}& \label{HModel:12}\mspace{18mu}\forall w,\mspace{-1mu}n\mspace{-1mu},\mspace{-1mu}i\mspace{-1mu},\mspace{-1mu}t;\\
&\textstyle{\mspace{9mu}D^{\text{m}}_{ni} I_{wnit} \leq D_{wnit} \leq D^{\text{M}}_{ni} I_{wnit},}
                                                                    \mspace{-180mu}& \label{HModel:13}\mspace{18mu}\forall w,\mspace{-1mu}n\mspace{-1mu},\mspace{-1mu}i\mspace{-1mu},\mspace{-1mu}t;
\end{flalign}
\end{subequations}

\subsubsection{Self-Scheduling Model With Estimated Water Values}\label{DAModel2}
Instead of using explicit storage targets, the medium-term planning could alternatively provide estimated water values $\hat{\nu}_{n}$\textemdash estimated net revenue that reservoir $n$ can yield in the future with an incremental unit of stored water. With this, the self-scheduling model can be formed as \eqref{SModel}, which is similar to \eqref{HModel} expect: \textit{i)} the objective function \eqref{SModel:1} includes an extra term of water value, and \textit{ii)} the target end-of-day
storage volume constraint \eqref{HModel:4} is removed.
\begin{subequations}\label{SModel}
\begin{flalign}
&\textstyle{\max\nolimits_{\Phi}}
\textstyle{\sum\limits_{t \in \mathcal{T}}G_{t}
+ p_{w}\sum\limits_{w \in \mathcal{W}}\sum\limits_{n \in \mathcal{N}}\hat{\nu}_{n} V_{w,n, T+1}}
- p_{w}\sum\limits_{w \in \mathcal{W}}\sum\limits_{t \in \mathcal{T}} C_{wt}
\mspace{-180mu}&\notag\\
&\text{where } 
 \Phi=\{\boldsymbol{C}, \boldsymbol{D}, \boldsymbol{G}, \boldsymbol{I}, \boldsymbol{P}, \boldsymbol{P}^{\text{act/cmit/vr}},
\boldsymbol{V}, \boldsymbol{W}^{\text{ws/i/o}}\}                                       \mspace{-180mu}&\label{SModel:1}\\
&\text{s.t. } \eqref{HModel:2}-\eqref{HModel:3}, \eqref{HModel:5}-\eqref{HModel:13};   \mspace{-180mu}&\label{SModel:2}
\end{flalign}
\end{subequations}

\vspace{-2.5mm}
\subsection{Real-Time Operation Model}\label{DetailRT}
\vspace{-1mm}
To closely follow the generation plans $P^{\text{sch}\star}_{t}$ during the RT stage, the RT operation model \eqref{RModel} is solved for each 5-minute SI $j_{t}$ of hour $t$ with actual VRES availability and natural water inflow. Its objective function \eqref{RModel:1} aims to make effective utilization of available water while closely following $P^{\text{sch}\star}_{t}$, of which the first term represents the penalty for under-generation and the second term rewards water conservation. Constraint \eqref{RModel:2} defines the under-generation; \eqref{RModel:3}-\eqref{RModel:6} are analogous to \eqref{HModel:7}-\eqref{HModel:10}; \eqref{RModel:7}-\eqref{RModel:9} resemble \eqref{HModel:11}-\eqref{HModel:13}. Parameter $\bar{W}_{nj_{t}}^{\text{RT,i}}$ in \eqref{HModel:6}, describing the amount of water discharged from upstream reservoirs in previous RT operation periods and injected into reservoir $n$ at SI $j_{t}$, takes the solutions of RT operation models in previous SIs.
\begin{subequations}\label{RModel}
\begin{flalign}
&\textstyle{\min\nolimits_{\Xi}
C^{\text{ud}}{P}^{\text{RT,}\Delta}_{j_{t}} - \sum\nolimits_{n \in \mathcal{N}}C^{\text{re}}
{V}^{\text{RT}}_{n, j_{t}+1}}                                        \mspace{-85mu}&                  \label{RModel:1}\\
&\text{where } \Xi=\{ \boldsymbol{D}^{\text{RT}}, \boldsymbol{I}^{\text{RT}},
\boldsymbol{P}^{\text{RT}}, P^{\text{RT,$\Delta$/act/vr}}_{j_{t}},
\boldsymbol{V}^{\text{RT}}, \boldsymbol{W}^{\text{RT,ws/o}}\}        \mspace{-85mu}&                            \notag\\
&\text{s.t.}\, \textstyle{P^{\text{RT,}\Delta}_{j_{t}}=P^{\text{sch}\star}_{t}-P_{j_{t}}^{\text{RT,act}},P^{\text{sch}\star}_{t}\geq P_{j_{t}}^{\text{RT,act}}};                       \mspace{-85mu}&                  \label{RModel:2}\\
&\mspace{20mu}\textstyle{P^{\text{RT,act}}_{j_{t}} = \sum\nolimits_{n \in \mathcal{N}} \sum\nolimits_{i \in \mathcal{I}_{n}} P^{\text{RT}}_{nij_{t}} + P^{\text{RT,vr}}_{j_{t}}};            \mspace{-85mu}&                  \label{RModel:3}\\
&\mspace{20mu}\textstyle{0 \leq P^{\text{RT,vr}}_{j_{t}}\leq \tilde{P}^{\text{vr}}_{j_{t}}};
                                                                     \mspace{-85mu}&                  \label{RModel:4}\\
&\mspace{20mu}\textstyle{V^{\text{m}}_{n} \leq V^{\text{RT}}_{n,j_{t}+1} \leq V^{\text{M}}_{n},}
                                                                      \mspace{-85mu}&\forall n;        \label{RModel:5}\\
&\mspace{20mu}\textstyle{V^{\text{RT}}_{n,j_{t}+1}=\bar{V}_{nj_{t}}^{\text{RT}}+ \tilde{W}_{nj_{t}}+\bar{W}^{\text{RT,i}}_{nj_{t}}-W^{\text{RT,o}}_{nj_{t}},}                           \mspace{-85mu}&                  \notag\\
&\mspace{20mu}\textstyle{W^{\text{RT,o}}_{nj_{t}} = \sum\nolimits_{i \in \mathcal{I}_{n}} \beta D^{\text{RT}}_{nij_{t}} + W^{\text{RT,ws}}_{nj_{t}},W^{\text{RT,ws}}_{nj_{t}} \geq 0,}         \mspace{-85mu}& \forall n;       \label{RModel:6}\\
&\mspace{20mu}\textstyle{P^{\text{RT}}_{nij_{t}}=\mathcal{F}^{\text{DtP}}(D^{\text{RT}}_{nij_{t}},I^{\text{RT}}_{nij_{t}}),}                                                                   \mspace{-85mu}&\forall n, i;     \label{RModel:7}\\
&\mspace{20mu}\textstyle{P^{\text{m}}_{ni} I^{\text{RT}}_{nij_{t}} \leq P^{\text{RT}}_{nij_{t}} \leq P^{\text{M}}_{ni} I^{\text{RT}}_{nij_{t}},}                                               \mspace{-85mu}&\forall n, i;    \label{RModel:8}\\
&\mspace{20mu}\textstyle{D^{\text{m}}_{ni} I^{\text{RT}}_{nij_{t}} \leq D^{\text{RT}}_{nij_{t}} \leq D^{\text{M}}_{ni} I^{\text{RT}}_{nij_{t}},I^{\text{RT}}_{nij_{t}} \in \{0, 1\},}                                               \mspace{-85mu}&\forall n, i;    \label{RModel:9}
\end{flalign}
\end{subequations}

Solving \eqref{RModel} reveals the under-generation $P^{\smash{\raisebox{-0.0ex}{\scriptsize RT,$\Delta$}}}_{j_{t}}$ for each SI $j_{t}$. If an under-generation occurs, the market participant is required to purchase back an equivalent amount of under-generation power at RT LMPs. However, self-scheduling producers, who already benefit from the privilege of high commitment priority, should bear a greater responsibility to meet their generation plans accommodated in the day-ahead market. To reflect this, a sufficiently large coefficient $C^{\text{ud}}$ is used in \eqref{HModel:6} and \eqref{RModel:1}.

\vspace{-2.5mm}
\bibliographystyle{IEEEtran}
\bibliography{Hydro_Refs_02}

\end{document}